\documentclass[12pt]{article}
\usepackage{eurosym}
\usepackage{amsfonts}
\usepackage{color,graphicx}
\usepackage{amsmath,amssymb,amsthm}
\usepackage{hyperref}
\usepackage{times}
\usepackage[letterpaper,hmarginratio=1:1,vmarginratio=1:1,heightrounded,tmargin=1in,lmargin=1in,footnotesep=.6cm]{geometry}
\usepackage{mathptmx}
\usepackage{accents}
\usepackage{graphicx}
\usepackage[style=nature]{biblatex}
\usepackage{booktabs}
\usepackage[labelfont=bf,format=plain]{caption}
\usepackage{rotating}
\usepackage{lscape}

\definecolor{Blue}{rgb}{0,0,0.8}

\graphicspath{{C:/Users/Ottaviani/Dropbox/clinicaltrials/paper/nature/}{C:/Dropbox/clinicaltrials/paper/latex/}}
\theoremstyle{plain}

\theoremstyle{definition}

\begin{document}

\author{J\'{e}r\^{o}me Adda\thanks{%
Department of Economics, BIDSA, and IGIER, Bocconi University, Via Roberto Sarfatti
25, 20136 Milan, Italy. Phone: +39--02--5836--5572. E-mail: \href{mailto:jerome.adda.work@gmail.com}%
{jerome.adda@unibocconi.it}.} \and Christian Decker\thanks{%
Department of Economics and UBS Center for Economics in Society, University of Zurich, Sch\"{o}nberggasse 1, 8001
Zurich, Switzerland. Phone: +41--44--634--61--26. E-mail: \href{mailto:christian.decker@econ.uzh.ch}%
{christian.decker@econ.uzh.ch}.} \and Marco Ottaviani\thanks{%
Department of Economics, BIDSA, and IGIER, Bocconi University, Via Roberto Sarfatti
25, 20136 Milan, Italy. Phone: +39--02--5836--3385. E-mail: \href{mailto:marco.ottaviani@unibocconi.it}%
{marco.ottaviani@unibocconi.it}.}}
\title{\vspace*{-1.15cm}
	\textbf{P-hacking in clinical trials and how incentives shape the distribution of results across phases\thanks{Funding by the European Research Council through grant 295835 (EVALIDEA) is
gratefully acknowledged. We thank Marco Bonetti, Tarani Chandola, Sylvain
Chassang, Francesco Decarolis, Edina Hot, John Ioannidis, Melissa Newham, Nicolas Serrano-Velarde, Tony Tse, and
Deborah Zarin for helpful comments. All authors have contributed equally. The authors declare no competing interests. A complete replication package is available upon request from the authors. This paper draws on Christian Decker's Master thesis ``P-Hacking in Clinical Trials?'', supervised by Marco Ottaviani and J\'{e}r\^{o}me Adda, and defended on April 20, 2017 at Bocconi University.} }}
\date{March 18, 2020}
\maketitle
\vspace*{-0.8cm}
\begin{abstract}
Clinical research should conform to high standards of ethical and scientific integrity, given that human lives are at stake. However, economic incentives can generate conflicts of interest for investigators, who may be inclined to withhold unfavorable results or even tamper with data in order to achieve desired outcomes. To shed light on the integrity of clinical trial results, this paper systematically analyzes the distribution of p-values of primary outcomes for phase II and phase III drug trials reported to the \textit{ClinicalTrials.gov} registry. First, we detect no bunching of results just above the classical 5\% threshold for statistical significance. Second, a density discontinuity test reveals an upward jump at the 5\% threshold for phase III results by small industry sponsors. Third, we document a larger fraction of significant results in phase III compared to phase II. Linking trials across phases, we find that early favorable results increase the likelihood of continuing into the next phase. Once we take into account this selective continuation, we can explain almost completely the excess of significant results in phase III for trials conducted by large industry sponsors. For small industry sponsors, instead, part of the excess remains unexplained.

\bigskip

\noindent\emph{Keywords:} Clinical trials; Drug development; Selective reporting; p-Hacking;
Economic incentives in research


\end{abstract}

\newpage
The evidence produced in clinical trials is susceptible to many kinds of bias \cite{Ioannidis2005,Garattini2016,Brown2018}. While some such biases
can occur accidentally, even unbeknownst to the study investigators, other biases may result from strategic
behavior of investigators and sponsors. In addition to the public value
of improving medical treatments, the information obtained through clinical
trials is privately valuable for the sponsoring pharmaceutical companies
that aim to demonstrate the safety and efficacy of newly developed drugs---the prerequisite for marketing approval by authorities such as the \textit{U.S. Food and Drug Administration} (FDA). Given the sizeable research and development costs involved \cite{DiMasi2003} and the lure of large potential profits, investigators can suffer from
conflicts of interest \cite{Relman1989,Angell2000,Lexchin2003,Budish2015} and pressure to withhold or ``beautify'' unfavorable results \cite{Boutron2018,Li2018} or even fabricate and falsify data
\cite{Fanelli2009}. 

In the 1990s and 2000s many medical
scholars began calling for more transparency in clinical research \cite{Young2008}, following public outcry over alarming evidence of
selective publication of trial results
\cite{Simes1986,Easterbrook1991,Turner2008}, cases of premature drug
approvals \cite{Rosen2007}, and allegations of data withholding
\cite{Harris2010}. As a response to these concerns, policymakers
established publicly accessible registries and result databases
\cite{Zarin2008,Zarin2017}, such as \textit{ClinicalTrials.gov}
\cite{Zarin2011,Tasneem2012} (see \hyperref[supp1:data]{\textit{SI Appendix}} for more details on the \textit{ClinicalTrials.gov} registry
and the legal requirements for reporting trial results).

\textit{ClinicalTrials.gov} now contains sufficient data to allow for a
systematic evaluation of the distribution of reported p-values. This is the
first such analysis, building on the literature that investigates
\textquotedblleft p-hacking\textquotedblright, publication bias, and the
\textquotedblleft file-drawer problem\textquotedblright  \cite{Rosenthal1979,Franco2014} for academic journal publications in a
number of fields, ranging from life sciences \cite{Holman2015} to psychology \cite{Simonsohn2014,Hartgernik2016}, political
science \cite{Gerber2008,Gerber2010}, and economics \cite{DeLong1992,Stanley2005,Brodeur2016}.

Given the escalation of stakes as research progresses through phases,
clinical trials are particularly well suited to detect how economic
incentives of sponsoring parties drive research activity \cite{Guedj2004,Krieger2017,Cunningham2019} and reporting bias.
Economic incentives in clinical trials may depend on the size of the sponsoring firm \cite{Guedj2004}. Compared to larger companies, smaller firms may have more to gain by misreporting results---and less reputation to lose if they are exposed. In other contexts, such reputational concerns have been found to vary by firm size \cite{Jin2009,Mayzlin2014} or by academic prominence \cite{Azoulay2017}. 

While the previous literature focused mostly on scientific publications in
academic journals for which pre-publication research results are typically
not observable, \textit{ClinicalTrials.gov} allows us to observe results
from clinical trials in earlier research phases. Thus, we are able to follow
the evolution of research results over time, and construct counterfactuals
not available in previous work. By linking trials across different phases of
clinical research, we are able to quantify the effect of the incentives to
selectively continue experimental research depending on early-stage results.

Our focus is on pre-approval interventional superiority studies on drugs
carried out as phase II and phase III trials. Trials in phase II investigate
drug safety and efficacy, typically with a small sample of experimental
subjects. Phase III trials investigate efficacy, while monitoring adverse
effects on a larger sample of individuals, and play a central role in
obtaining approval to market the drug from regulators such as the FDA. To facilitate the analysis, we transform the
p-values into test statistics, supposing that they would all originate
from a two-sided Z-test of a null hypothesis that the drug has the same
effect as the comparison. This transformation allows us to investigate both
the overall shape of the distribution and the region around the thresholds
for statistical significance more easily 
(see \hyperref[MMpz]{Materials and Methods} and \hyperref[supp1:pz]{\textit{SI Appendix}} for further information on the data and
the p-z transformation).

\subsection*{The Distribution of z-Scores: Irregularity Tests}

\begin{figure}[tbp]
\centering
\includegraphics[width=\linewidth]{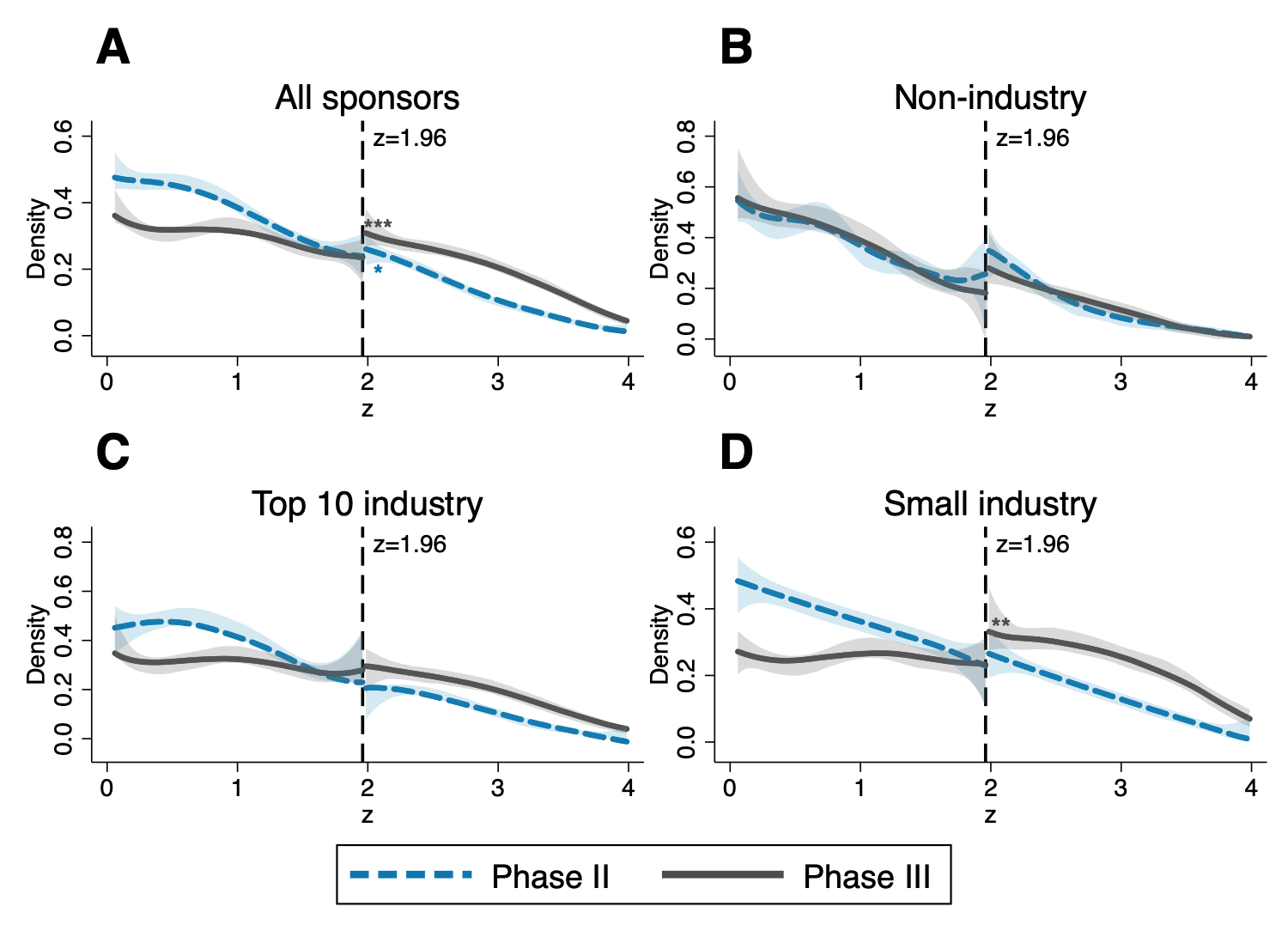}
\caption{\textbf{Comparison of phase II and phase III densities of the z-score
and tests for discontinuity at $z=1.96$, depending on affiliation of lead
sponsor.} Density estimates of the constructed z-statistics for primary
outcomes of phase II (dashed blue lines) and phase III (solid grey lines)
trials. The shaded areas are 95\%-confidence bands and the vertical lines at
1.96 correspond to the threshold for statistical significance at 0.05 level.
Sample sizes: A: $n=3,953$ (phase II), $n=3,664$ (phase III); B: $n=1,171$ (phase II), $n=720$ (phase III); C: $n=1,332$ (phase II), $n=1,424$ (phase III); D: $n=1,450$ (phase II), $n=1,520$ (phase III). Significance levels for discontinuity tests 
\cite{Cattaneo2019}: * $p<0.1$; ** $p<0.05$; *** $p<0.01$; exact p-values reported in \hyperref[tab:cattaneo1]{Table \ref*{tab:cattaneo1}}.}
\label{fig:discontinuities}
\end{figure}

\hyperref[fig:discontinuities]{Figure \ref*{fig:discontinuities}} displays
density estimates of the constructed z-statistics for tests performed for
primary outcomes of phase II and phase III trials. We present results for all
trials in panel A and subsequently provide the break down by affiliation of the lead
sponsor: non-industry (NIH, US federal agencies, universities,
etc.) in panel B, top ten industry (the ten pharmaceutical companies in the
sample with largest revenues in 2018; see \hyperref[tab:rankings]{Table \ref*{tab:rankings}})
in panel C, and small industry (the
remaining smaller pharmaceutical companies) in panel D.

Next, we diagnose three possible irregularities in the distribution of
z-statistics of trials, at or above the 5\% significance threshold,
corresponding to a z-statistic of 1.96. Further technical details and robustness checks are gathered in the \hyperref[supp2]{\textit{SI Appendix}}.

\begin{enumerate}
	\item \textbf{Spike in the density function just above 1.96.}\\
	We detect no spikes in the densities (or discontinuities in the
	distribution functions) just above 1.96, the salient significance
	threshold. Such spikes, indicating that results are inflated to clear the
	significance hurdle, have been documented in previous studies of
	z-distributions for tests in academic publications across life sciences \cite{Holman2015} as well as economics \cite{Brodeur2016} and
	business studies \cite{Meyer2017}. Thus, the more natural distribution of z-scores
	from \textit{ClinicalTrials.gov} displays more integrity
	compared to results reported for publications in scientific journals. This difference may partially be explained by the absence of the additional layer of editorial selection, which may be based also on the statistical significance of presented results. This first finding suggests that registered results are not inflated at the margin just to clear the significance threshold.
	
	\item \textbf{Discontinuity of the density function at 1.96.}\\
	We investigate the presence of a discontinuity in the density of
	z-statistics with a test that relies on a simple local polynomial density
	estimator \cite{Cattaneo2019}. The densities for phase II trials are
	smooth and do not show a noteworthy upward shift at the 1.96 threshold in
	all cases. In contrast, the densities of z-statistics for industry-sponsored
	(both small and top ten) phase III trials display a break at 1.96. The break
	is statistically significant only for phase III trials undertaken by small
	pharmaceutical companies (panel D), with a persistent upward shift to the
	right of the threshold, indicating an abnormal amount of significant
	results. This pattern is
	suggestive of \textit{selective reporting}, i.e., strategic concealment of
	some non-significant results.
	
	The different patterns observed between large and small industry sponsors (Panels C and D) are robust across a wide range of alternative ways to define ``large'' sponsors (\hyperref[fig:rob_disc]{Figure \ref{fig:rob_disc}}). Moreover, we find a similar discontinuity for phase III trials by small industry sponsors when transforming p-values to test statistics of a one-sided instead of a two-sided test (\hyperref[fig:discontinuities1s]{Figure \ref{fig:discontinuities1s}}).
	
	\item \textbf{Excess of significant results in phase III compared to phase II.}\\
	\hyperref[fig:discontinuities]{Figure \ref{fig:discontinuities}}
	indicates an excess of favorable results over the 1.96 threshold in
	phase III compared to phase II. More
	favorable results are more likely to be observed in phase III than in phase II. The phase III distribution of
	z-statistics stochastically dominates the phase II distribution. Dominance is particularly strong for industry-sponsored trials (Panels C and
	D). This pattern appears suspicious, but it is not as alarming as a spike at the significance threshold. Whilst only 34.7\% of phase II trial
	results by non-industry sponsors fall above 1.96 (and 34.8\% respectively
	for phase III, a difference that is not statistically significant), the fraction of significant results rises to
	45.7\% in phase II and 70.6\% in phase III for industry-sponsored trials.
\end{enumerate}

Recall that the analysis above considers only p-values associated to primary
outcomes of trials. These results constitute the main measure for success of
the treatment being trialed, for both the investigators themselves and the
evaluating authorities. The densities of z-scores from lower-stake secondary outcomes for all groups of sponsors and both phases do not display any meaningful discontinuity at the significance threshold (see \hyperref[fig:discontinuities_sec]{Figure \ref*{fig:discontinuities_sec}} and \hyperref[tab:cattaneo1_sec]{Table \ref*{tab:cattaneo1_sec}}). Moreover, for secondary outcomes the excess of significant results
from industry-sponsored trials in phase III relative to phase II is much
smaller compared to the distribution for primary outcomes. We find
irregularities only for higher-stake primary outcomes, suggesting that
incentives of reporting parties play a role.

\subsection*{Linking Trials across Phases: Controlling for Selective Continuation}

The FDA focuses mainly on phase III results when deciding about marketing
approval, a decision with major financial consequences for pharmaceutical
companies. Given these incentives, the observed excess of significant
results particularly in the group of industry-sponsored phase III trials could
be interpreted as evidence of tampering (\textit{p-hacking}) or
non-disclosure of negative results (\textit{selective reporting}). However,
this conclusion would be premature without first carefully examining the
dynamic incentives underlying clinical research, as we set out to do.

An alternative explanation for the excess of significant results in
phase III relative to phase II is the \textit{selective continuation} of drug
testing to the next phase only when initial results are sufficiently
encouraging. \textit{Selective continuation} saves on costly clinical research and can
thus even be socially desirable, as long as such economic considerations do not distort research activity away from important but costly projects \cite{Budish2015}. Also, from an ethical viewpoint, no further
trials with volunteer patients should be conducted when a drug is highly
unlikely to have a positive impact. Time and resources should be devoted to more promising projects instead. We outline a model of the sponsor's continuation decision in \hyperref[supp4:selection]{Materials and Methods}.

\begin{figure}[t]
	\centering
	\includegraphics[width=.8\linewidth]{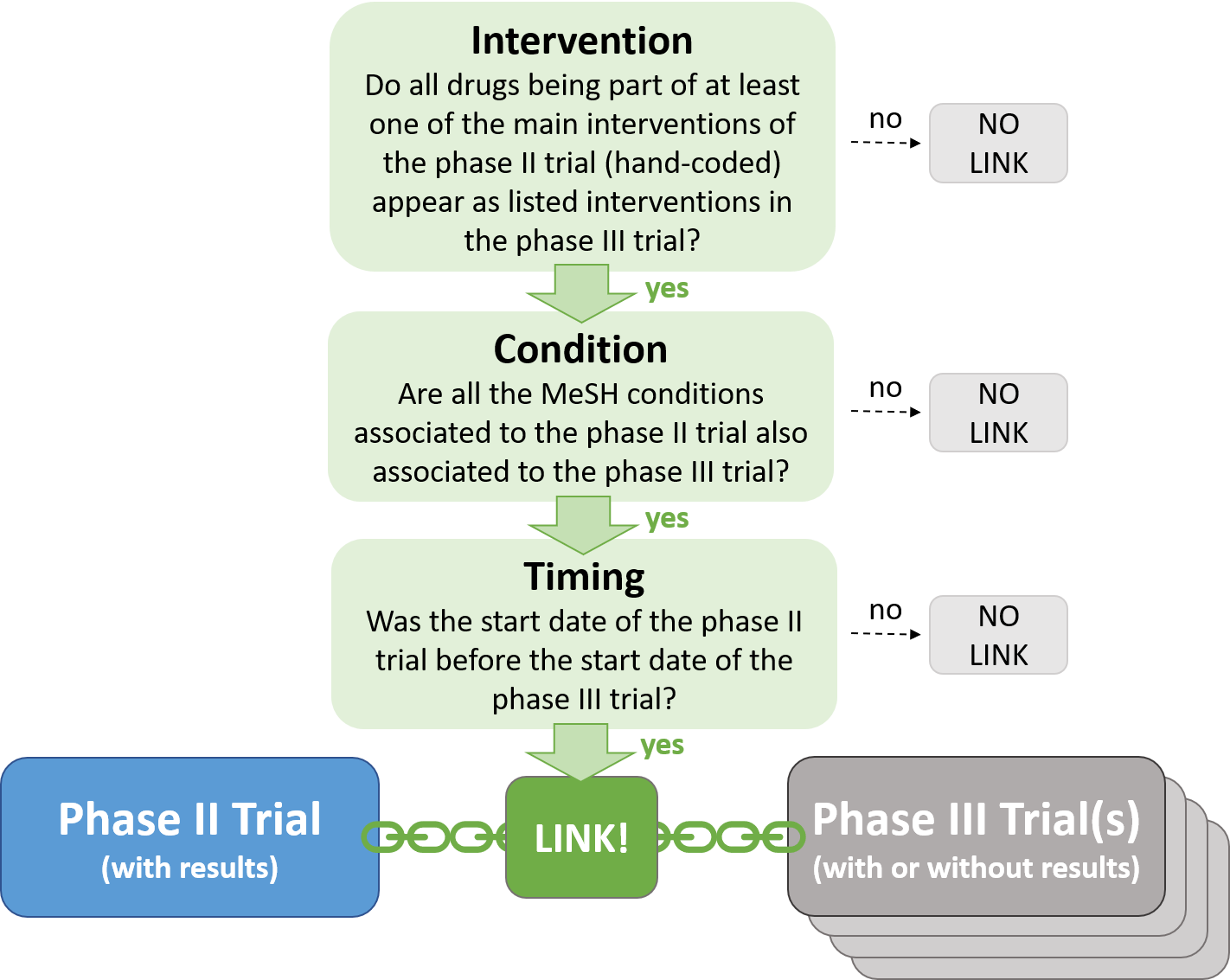}
	\caption{\textbf{Linking phase II and phase III trials.} We consider a phase II trial as continued if we found at least one phase III trial registered in the database (regardless of whether associated results are reported or not) fulfilling all three criteria (intervention, condition, and timing). See \hyperref[supp3]{\textit{SI Appendix}} for a more detailed description of the linking procedure.}
	\label{fig:linking}
\end{figure}

To identify the impact of \textit{selective continuation}, we develop a
procedure to link phase II and phase III trials in our dataset based on the
main intervention (i.e., the tested drug or combination of drugs), the medical
condition to be treated, and the timing. This procedure is illustrated in \hyperref[fig:linking]{Figure \ref*{fig:linking}}. A given phase II trial may either (i) have no corresponding phase III trial with the same
intervention and same condition, or (ii) have one or multiple matches in
phase III. In the latter case, we consider the phase II trial as continued into phase III. The resulting linked data, which we make available to the research
community, is a key input in the methodology we develop to estimate a
selection function capturing \textit{selective continuation} for
industry-sponsored trials.

Following our model of the firm's continuation decision, we estimate the selection function with a logistic regression of a dummy
variable indicating if there is at least one match among the phase III trials
in the database (regardless of whether phase III results are reported or not)
on the phase II z-score. We control for adjustment for multiple hypothesis
testing, a flexible time trend, and other covariates that might influence the
perceived persuasiveness of phase II results (square root of overall
enrollment to each trial as proxy for power of the statistical tests, active
comparator vs. placebo) or the economic incentives to undertake research
(fixed effects for the treated condition) on top of the z-score; see \hyperref[supp4:selection]%
{Materials and Methods} for the exact specification. The predicted values of this
selection function can be interpreted as the probability that a drug progresses to
phase III conditional on the information available at the end of phase II, consisting of the phase II z-score and other covariates. 

In most cases, very low p-values are no longer reported precisely but only
as being below the thresholds 0.001 or 0.0001 (e.g. $p<0.001$ instead of $%
p=0.0008$). Therefore, we estimate the continuation probability separately
for those two cases by including dummies for ``$z>3.29$'' (corresponding to
the p-value being reported as $p<0.001$) and ``$z>3.89$'' (corresponding to $%
p<0.0001$) in the specification of the selection function. 

\begin{table*}
	\caption{\textbf{Estimates of logit selection function for \textit{selective
				continuation}, based on primary outcomes.}}
	\label{tab:selection}
	\centering
	
	\begin{tabular}{lccc}
		\toprule
		& \textbf{(1)} & \textbf{(2)} & \textbf{(3)} \\
		\textbf{Sponsor} & \textbf{All} & \textbf{Small} & \textbf{Top 10} \\
		& \textbf{industry} & \textbf{industry} & \textbf{industry} \\
		\midrule
		&       &       &  \\
		Phase II z-score & 0.331*** & 0.266*** & 0.404*** \\
		& (0.0793) & (0.100) & (0.130) \\
		Dummy for phase II z-score reported as ``$z>3.29$'' & 1.063*** & 0.756** & 1.750*** \\
		& (0.226) & (0.329) & (0.373) \\
		Dummy for phase II z-score reported as ``$z>3.89$'' & 1.232*** & 0.787*** & 1.643*** \\
		& (0.255) & (0.285) & (0.446) \\
		Mean dependent variable & 0.296 & 0.344 & 0.246 \\
		P-value Wald test (2)=(3) &       & \multicolumn{2}{c}{0.00480} \\
		&       &       &  \\
		Controls & yes   & yes   & yes \\
		MeSH condition fixed effects & yes   & yes   & yes \\
		Completion year fixed effects & yes   & yes   & yes \\
		Observations & 3,925 & 2,017 & 1,908 \\
		No. of trials & 1,167 & 674   & 493 \\
		\bottomrule
	\end{tabular}%
	
	\caption*{\small \textit{Notes:} Unit of observation: trial-outcome;
		included controls: square root of the overall enrollment, dummy for placebo
		comparator, and dummy for multiple hypothesis testing
		adjustment. See \hyperref[supp4:selection]{Materials and Methods} for the exact specification. Categories for condition fixed effects are based on Medical Subject Headings (MeSH) terms associated to the trials \cite{Tasneem2012}; for more details, see \hyperref[supp1:mkt]{\textit{SI Appendix}}. ``P-value Wald test (2)=(3)'' reports the p-value of a Wald test of the null hypothesis of joint equality of the coefficients in the first three rows and the constant between column 2 and column 3. Standard errors in parentheses are
		clustered at the MeSH condition level; significance levels (based on a two-sided
		t-test): * $p<0.1$; ** $p<0.05$; *** $p<0.01$. }
\end{table*}

\begin{figure}
	\centering
	\includegraphics[width=.8\linewidth]{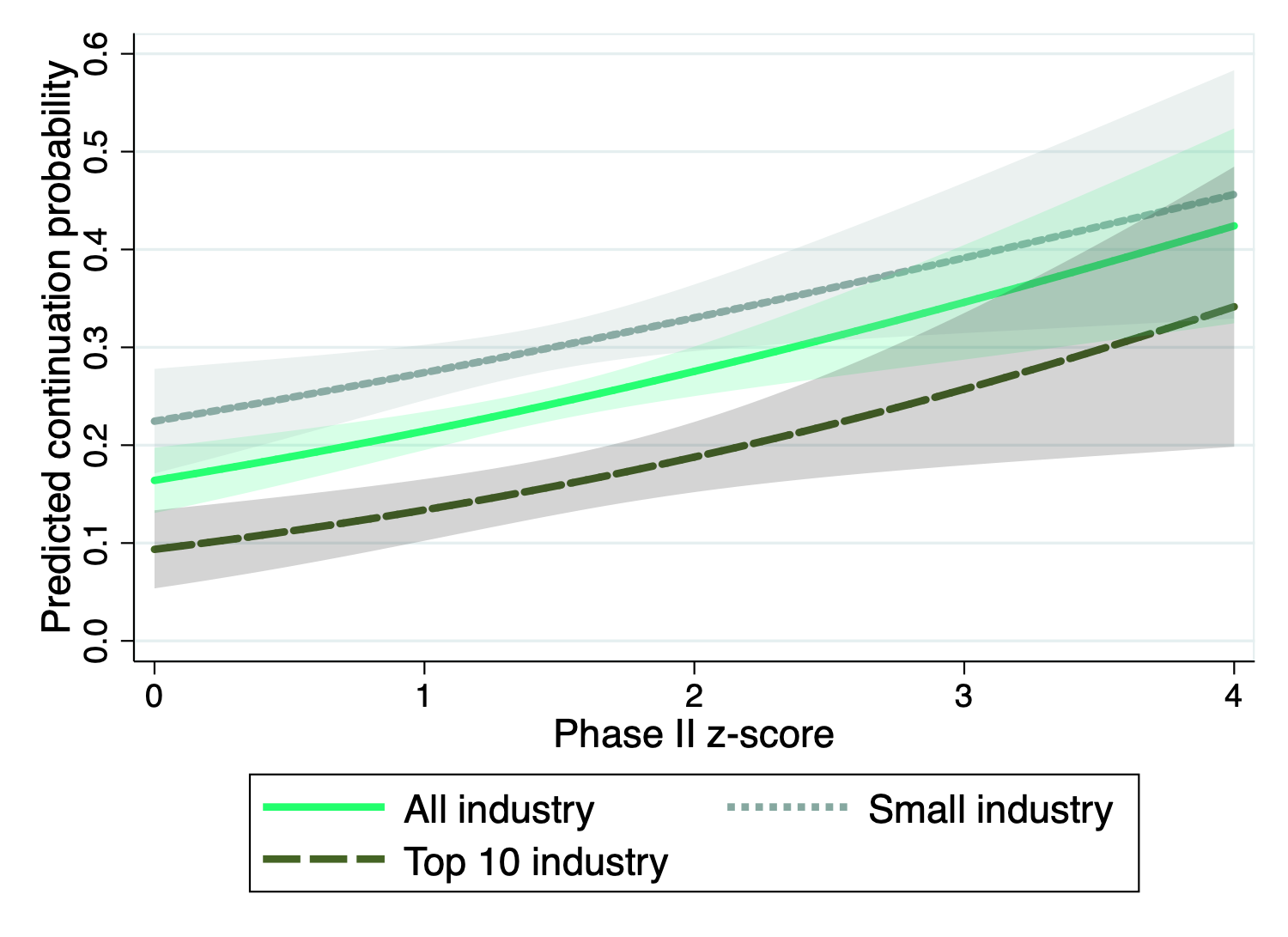}
	\caption{\textbf{Predicted continuation probability as function on the phase II z-score, depending on affiliation of lead sponsor.} Predictions are based on the estimated logit selection functions for \textit{selective continuation};  see \hyperref[tab:selection]{Table \ref*{tab:selection}} for the estimated coefficients. All control variables
		are fixed at their mean values. The shaded areas are 95\%-confidence bands.}
	\label{fig:selection}
\end{figure}

\hyperref[tab:selection]{Table \ref*{tab:selection}} displays the estimated logit coefficients for all industry sponsors (column 1), and for small and Top 10 industry sponsors separately (columns 2 and 3, respectively). \hyperref[fig:selection]{Figure \ref*{fig:selection}} illustrates the estimated selection functions graphically.
The solid green line shows the predicted continuation probability as function of the phase II z-score.
A higher z-score in phase II significantly increases the probability of continuation
to phase III. The lighter dotted and darker dashed lines show the predictions when considering only trials
conducted by small sponsors or respectively the ten largest industry sponsors. The
estimated continuation probabilities suggest that larger companies continue research projects more
selectively. The overall share
of matched trials is lower for large industry sponsors, captured by the downward shift of the selection function.

In the context of our model of the firm's continuation decision, the continuation probability is negatively associated with the opportunity cost of continuing a specific project. On average, this cost can be expected to be greater for large sponsors with many alternative projects. This interpretation is in line with findings from previous studies arguing that managers of larger firms with multiple products in development have less private costs attached to terminating unpromising research projects and thus are more efficient \cite{Guedj2004}.

In \hyperref[tab:selection_sec]{Table \ref*{tab:selection_sec}}
we report estimates of the same logistic model when considering the phase II
z-scores associated to secondary outcomes instead of primary outcomes.
The coefficients related to the z-score are much smaller in magnitude, and most of the coefficients are not statistically significant, notwithstanding the much larger sample size. This finding confirms that the
evaluation of a trial's success, and therefore also \textit{selective
	continuation}, is based predominantly on primary outcomes.

\subsection*{Decomposition of the Difference in Significant Results between
	Phase II and Phase III}

Under the assumption that, conditional on our control variables, the
expected z-statistic in phase III equals the z of a similar phase II trial, we
can construct a hypothetical phase III distribution for primary outcomes
accounting for \textit{selective continuation}. To do so, we estimate the
kernel density of phase II statistics (for now disregarding \textquotedblleft 
$z>3.29$\textquotedblright\ and \textquotedblleft $z>3.89$\textquotedblright
) reweighting each observation by the \textit{continuation} probability
predicted by our selection function given the characteristics of the phase II
trial. The resulting counterfactual density can be compared to the actual
phase II and phase III densities which we estimate using a standard unweighted
kernel estimator.

Since the selection function is increasing in the phase II z-score, the
counterfactual z-density rotates counter-clockwise, increasing the share of
significant results (see \hyperref[fig:densities]{Figure \ref*{fig:densities}}). To calculate the overall share of significant results under the
hypothetical regime, we combine the estimated densities with the number of
\textquotedblleft $z>3.29$\textquotedblright\ and \textquotedblleft $z>3.89$%
\textquotedblright\ results predicted from the selection functions and
renormalize to one.

\begin{figure}[p]
\centering
\includegraphics[width=\linewidth]{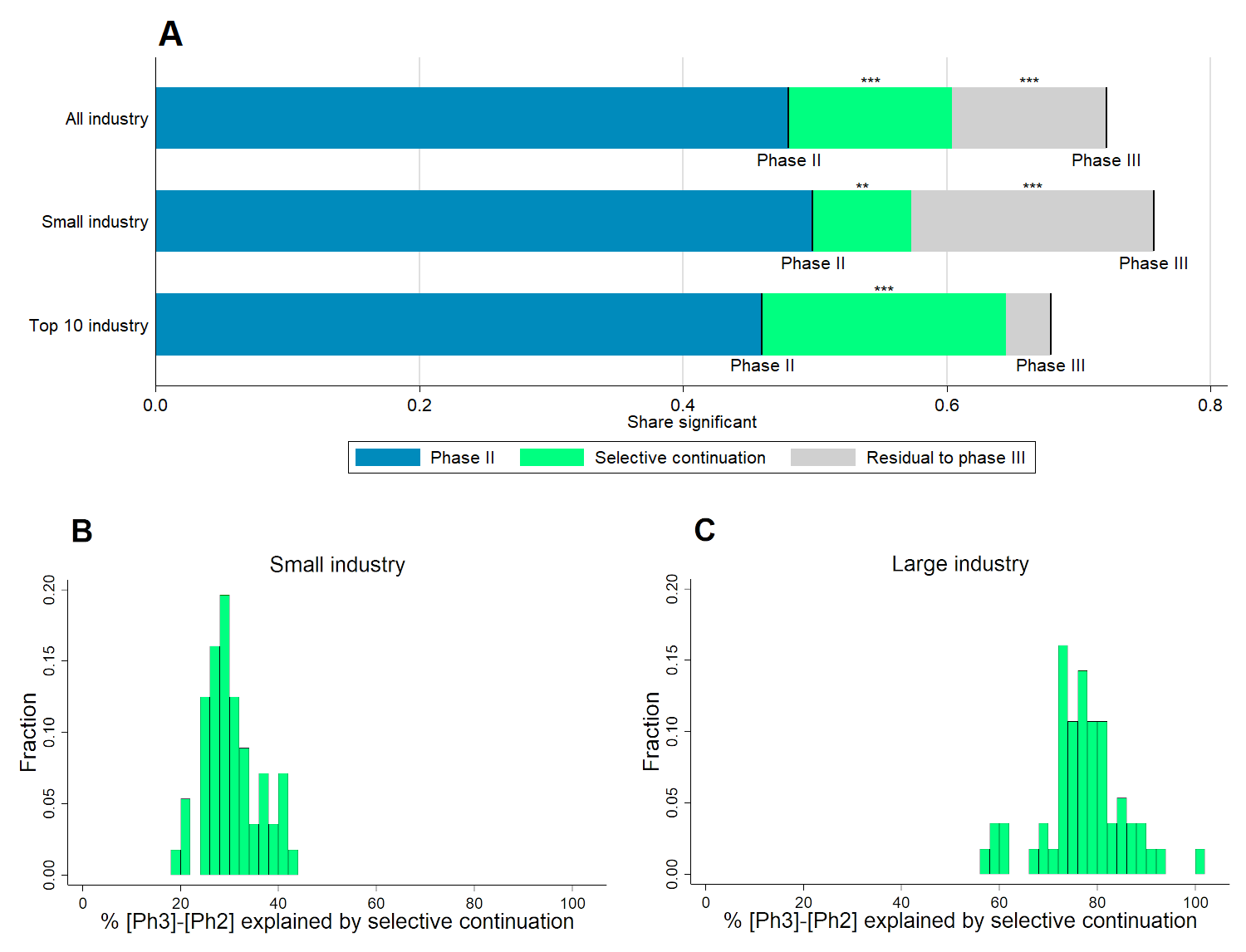}
\caption{\textbf{Panel A: Selection-based decomposition of the difference in
		significant results from primary outcomes between phase II and phase III,
		depending on affiliation of lead sponsor (top ten revenues criterion).} Phase II and III lines represent the
	shares of trials with a p-value below 5 percent (or equivalently a z-score
	above 1.96). The green segments represent the parts of the differences
	explained by \textit{selective continuation}, based on counterfactuals
	constructed from the phase II distribution. For precise numbers and sample
	sizes see \hyperref[tab:decomposition]{Table \ref*{tab:decomposition}}. Significance levels for the differences (based on a
	two-sided t-test): * $p<0.1$; ** $p<0.05$; *** $p<0.01$.
	\textbf{Panel B and C: Histograms of the percentage share of the difference in the share of significant results between phase III and phase II explained by \textit{selective continuation} across different definitions for large vs. small industry sponsors.} The shares correspond to the green area in panel A divided by the sum of the green and the grey areas. The sample of industry sponsored trials is split according to 56 different definitions of large sponsors. These definitions are obtained by ranking sponsors by their 2018 revenues, volume of prescription drug sales in 2018, R\&D spending in 2018, and the number of trials reported to the registry. For each of these four criteria, 14 different definitions of ``large vs. small'' are created: top seven vs. remainder, top eight vs. remainder, and so on up to top twenty vs. remainder. Further details are provided in \hyperref[supp1:pz]{\textit{SI Appendix}}.}
\label{fig:decomposition}
\end{figure}

Based on this construction, we decompose the difference in the share of
significant results in phase II and phase III into two parts: \textit{selective
	continuation} and an unexplained residual. As illustrated in \hyperref[fig:decomposition]%
{Figure \ref*{fig:decomposition}}, panel A and \hyperref[tab:decomposition]{Table \ref*{tab:decomposition}}, when we consider all industry
sponsored trials, \textit{selective continuation}, i.e., economizing on the
cost of trials that are not promising enough, accounts for more than half of
the difference, leaving 48.5\% of the difference unexplained.

Next, we repeat the estimation procedure separately for trials sponsored by
large and small industry. The difference in the share of significant results
between phase II and phase III is slightly larger for trials by small sponsors (21.9 percentage
points for top ten industry vs. 25.8 percentage points for small industry).
For trials sponsored by the ten largest companies, the difference between the actual share of significant phase III results and the share predicted by \textit{selective continuation} from phase II shrinks to 3.4 percentage
points and is no longer statistically significant.
Thus, for top ten industry sponsors our methodology suggests no indication of \textit{selective reporting} or potential tampering: \textit{selective continuation} can explain almost the entire excess share of significant results in phase III trials compared to phase II trials.

A different picture emerges for small industry sponsors. According to the selection function estimated in \hyperref[tab:selection]{Table \ref*{tab:selection}} and displayed in  \hyperref[fig:selection]{Figure \ref*{fig:selection}}, small sponsors are much more likely to proceed to phase III than large sponsors, especially following phase II trials with relatively low z-statistics. Hence, for small sponsors \textit{selective continuation} is less pronounced and can only account for less than one third of the excess share of significant results in phase III trials compared to phase II trials. Phase III results actually reported by small sponsors appear to be much more favorable than predicted by the selection function; for these sponsors we are left with a statistically significant unexplained residual of 18.4 percentage points, as displayed in \hyperref[fig:decomposition]{Figure \ref*{fig:decomposition}}, panel A. 

As illustrated by \hyperref[fig:decomposition]%
{Figure \ref*{fig:decomposition}}, panel B and C, these different patterns between large and small industry sponsors are robust across a wide range of alternative ways to define ``large'' sponsors. For small sponsors (panel B), the share of the explained difference ranges between 19\% and 44\% with the majority of results being very close to the estimate in our main specification (29\%). Also for different definitions of large sponsors (panel C), the estimates are quite close to the result from our main specification (85\%), ranging between 57\% and 101\%.

These findings are consistent with our earlier observation that small industry is the only group of sponsors for which the phase III z-density exhibits a statistically significant discontinuity at the 1.96 threshold. Along the same lines, a recent evaluation of compliance with FDA requirements for reporting of trial results to \textit{ClinicalTrials.gov} finds that compliance improves with sponsor size \cite{DeVito2020}.

\subsection*{Discussion and Conclusion}

Overall, the distribution of z-scores from \textit{ClinicalTrials.gov} does
not indicate widespread manipulation of results reported to the registry. Given the increasing adoption of randomized control trials across life and social sciences, our findings speak in favor of setting up repositories similar to \textit{ClinicalTrials.gov} in these other domains to monitor results and improve the credibility of research.

As we show, to correctly interpret the distribution of research results, it is important to understand the sequential nature of research and its interplay with economic incentives. 
Although phase III trials appear to deliver too many positive results, we can explain a large part of this excess of favorable results by linking them to phase II outcomes and accounting for \emph{selective continuation}. 

However, we find that \emph{selective continuation} cannot explain fully the high number of significant results in phase III trials sponsored by smaller firms. For the same group of trials, we also identify a discontinuity in the density at the classical significance threshold.  
These patterns suggest that enforcers of registration should pay particular attention to smaller industry sponsors, for which reputational concerns may be less consequential---a channel that should be investigated more thoroughly by future work. 

In conclusion, our exploratory findings indicate that current levels of regulation and enforcement are not sufficient to fully discipline reporting. To evaluate opportunities for reform, policy makers might want to weigh the ex post information benefits of mandatory registration against the reduced incentives of investigators to undertake clinical trials \cite{Matthews1985,Dahm2009,Henry2009,Polinsky2012,Henry2019}. An empirical quantification of this chilling effect could serve as an important input for a social cost-benefit analysis for tightening current rules.

%
%

%
%

\clearpage

\section*{Materials and Methods}

\subsection*{\textit{Database for Aggregate Analysis of ClinicalTrials.gov} (AACT)}

The \textit{Database for Aggregate Analysis of ClinicalTrials.gov} (AACT)
was launched in September 2010 to allow for free bulk download of all the data contained in the \textit{ClinicalTrials.gov} registry \cite{Zarin2017, Zarin2011,
	Tasneem2012}. The project is administered by the 
\textit{Clinical Trials Transformation Initiative} (CTTI), a partnership of
the FDA and Duke University with the aim of improving quality and efficiency
of clinical trials. The database, which is updated daily and directly
accessible in the cloud, contains over 40 sub-tables with information on timing,
conditions, interventions, facilities, locations, sponsors, investigators,
responsible authorities, eligible participants, outcome measures, adverse
events, results, and descriptions of trials. 

The trials in the database cover a wide range of different diseases,
interventions, and study designs. Hence, also the reported results are very
diverse in nature. In contrast to a meta-analysis on a specific disease or
treatment, which typically uses only a narrowly defined subgroup of the
dataset, we analyze the largest possible portion of the overall data. Given the aggregate level of our analysis, rather than using the
estimated coefficients, we focus on p-values, the only measures reported
uniformly and comparably for many trials, independent of their
characteristics and the statistical method used for the analysis.

This study is based on the AACT data available on August 15, 2019. Over the last two years, we obtained similar results in earlier drafts of this paper based on less data. We
concentrate on phase II and phase III interventional (as opposed to
observational) superiority (as opposed to non-inferiority) studies on drugs
(as opposed to medical devices and others) which report at least one proper
p-value for a statistical test on a primary outcome of the trial.

We drop the trials of the sponsor \textit{Colgate Palmolive}, which
reported p-values exactly equal to 0.05 for 137 out of its 150 results. We
attribute these exact p-values of 0.05 to a reporting mistake; clearly these
were intended to be reported as significant results with p-value lower or
equal to 0.05. Leaving \textit{Colgate Palmolive}'s results in the sample
would lead to a substantial spike at $z=1.96$ which could be wrongly
interpreted as evidence for p-hacking. Moreover, we drop the trial with the identifier \textit{NCT02799472} as it reports 211 p-values for primary outcomes and would therefore have much more impact than all other trials (average number of p-values for primary outcomes per trial: 2.5, median: 1).

Altogether, we obtain a sample of 12,621 p-values from tests performed on
primary outcomes of 4,977 trials. These single p-values constitute the units
of observation for our analysis. As a consequence of
the \textit{FDA Amendments Act} (FDAAA), the largest part of our results
data pertains to trials conducted after 2007.

\subsection*{p-z Transformation}
\label{MMpz}
We transform the p-values taken from the AACT database to corresponding z-statistics by supposing that all
p-values would originate from a two-sided Z-test of a null hypothesis that
the drug has the same effect as the comparison. Given that under the null
hypothesis this statistic is normally distributed, we have the one-to-one
correspondence $z=-\Phi ^{-1}(\frac{p}{2})$, where $z$ is the absolute value
of the test-statistic and $\Phi ^{-1}$ is the inverse of the standard normal
cumulative distribution function. This transformation facilitates both the graphical analysis and the identification of discontinuities at the significance threshold, given that the z-density is close to linear around the significance threshold, whereas the corresponding p-density is highly nonlinear in this range.

\subsection*{Density Discontinuity Tests}
We implement tests of discontinuity in the z-score
density at the $z=1.96$ significance threshold based on the state-of-the-art procedure developed by Cattaneo, Jansson,
and Ma \cite{Cattaneo2019}. This test builds on a local polynomial
density estimation technique that avoids pre-binning of the data. More details on the testing procedure and supplementary results can be found in \hyperref[supp2]{\textit{SI Appendix}}.

\subsection*{Linking Phase II and Phase III Trials}

To analyze \textit{selective continuation} from phase II to
phase III, we link phase II and phase III trials in our dataset, based on the main intervention, the medical condition to be treated, and the timing.

We read one by one the protocols for all the phase II trials in the dataset for which at least one p-value is
reported. We consider only phase II trials that were completed before end of December 2018 to allow for enough time such that a follow-up phase III trial could have been registered by August 2019. From the protocols, we determine the main
experimental intervention(s), i.e., the main drug or combination of drugs
whose efficacy and safety is to be established, for 1,773 phase II trials.

We consider a phase II trial as continued if we could link it to at least one phase III trial. That is, if we found at least one phase III trial registered in the
database (regardless of whether associated results are reported or not)
fulfilling all of the following criteria:

\begin{enumerate}
	\item \textbf{Intervention:} All drugs being part of at least one of the determined main
	interventions of the phase II trial appear as listed interventions in the
	phase III trial. This is either with exactly the same name or with a synonym
	which the reporting party states to refer to the same drug.
	
	\item\textbf{Condition:} All the MeSH-conditions \cite{Tasneem2012} associated to the phase II trial are also
	associated to the phase III trial.
	
	\item \textbf{Timing:} The start date of the phase II trial was before the start date of the
	phase III trial.
\end{enumerate}
For more details on the linking procedure, see \hyperref[supp3]{\textit{SI Appendix}}.

\subsection*{Selection Function}

\label{supp4:selection}

Denote by $I_2$ a vector collecting the relevant information pertaining to the clinical trial at the end of phase II. It contains the z-score, $z^{Ph2}$, and other variables describing the circumstances of the trial (such as sample size to proxy for statistical power). If the sponsor firm decides to stop the development of the drug, it obtains a payoff of $\underline{V}(I_2)+\underline{\eta}$. In case of continuation into phase III, the firm pays a development cost $c+\eta$. The idiosyncratic payoff and cost shocks $\underline{\eta}$ and $\eta$ are only observable to the firm, but not to the econometrician. The future payoff is denoted $V^{Ph3}$ and is increasing in the phase III z-score, which is uncertain at the time of the decision to set up a phase III trial. The firm has an expectation on the distribution of the z-score, based on the information available in $I_2$. The decision of the firm is thus
\begin{align*}
V^{Ph2}(I_2)=\max\left[\underline{V}(I_2)+\underline{\eta}; -c-\eta+\delta E_{z_3|I_2}V^{Ph3}(z_3) \right],
\end{align*}
where $\delta$ is the discount factor. Assuming that the idiosyncratic shocks $\underline{\eta}$ and $\eta$ are both iid and extreme value distributed, the probability of undertaking a phase III trial is a logistic function \cite{McFadden1978}
\begin{align*}
Prob(continuation)&=\frac{\exp( -c+\delta E_{z_3|I_2}V^{Ph3}(z_3))}{exp(\underline{V}(I_2))+\exp( -c+\delta E_{z_3|I_2}V^{Ph3}(z_3))} \\
&=logistic(I_2).
\end{align*}

Following this model, we use a logistic regression to estimate a selection function that captures \textit{selective continuation} for industry-sponsored trials. In the sample of phase II z-scores, restricted as explained in the section above, we estimate the logistic model 
\begin{align*}
continuation_{i}=logistic\left[ \alpha +\beta
_{0}(1-D1_{i}-D2_{i})z_{i}^{Ph2}+\beta _{1}D1_{i}+\beta _{2}D2_{i}+\mathbf{x}%
_{i}^{\prime }\mathbf{\gamma }+\phi_{ci}+\tau_{ti}+\varepsilon _{i}\right] \mbox{,}
\end{align*}%
where $continuation_{i}$ is a dummy variable which results from our linking of
trials across phases and equals one if there is at least one phase III trial
matched to phase II trial to which z-score $i$ belongs (regardless of whether
results are reported), and $z_{i}^{Ph2}$ is the phase II z-score associated
to a primary outcome. $D1_{i}$ and $D2_{i}$ are dummy variables for a
statistic to be reported as \textquotedblleft $z>3.29$\textquotedblright\ or
\textquotedblleft $z>3.89$\textquotedblright , respectively. As explained
above, those cases are so frequent that we treat them separately.

Moreover, the vector $\mathbf{x}_{i}$ gathers further control variables
which might influence the perceived persuasiveness of phase II results or the
economic incentives to carry on with the research on top of the z-score.
These include the square root of the overall enrollment to each trial (as
proxy for the power of the tests), a dummy indicating whether there was a
placebo involved in the trial (as opposed to an active comparator), and a dummy indicating whether the
p-value is explicitly declared as adjusted for multiple hypothesis testing.
For the last variable, the baseline corresponds to no adjustment of the
critical value of the testing procedure or no information provided. We
codified this variable manually from the p-value descriptions; only 2.9\% of
the relevant observations are explicitly adjusted.

To account for potential systematic differences across drugs for the treatment of different kinds of conditions, we include condition fixed effects $\phi_c$. For this purpose, we assign each trial in one of the 15 largest categories of conditions, based on the MeSH terms determined by the curators of the database \cite{Tasneem2012}. For more details, see \hyperref[supp1:mkt]{\textit{SI Appendix}}.

As registration of trials and reporting of results occurs often with a
substantial time lag, we also control for a flexible time trend by including completion year fixed effects $\tau_t$.

Summing up, $z^{Ph2},D_1,D_2$, $\mathbf{x}$, and $\phi_c$ correspond to $I_2$, the information relevant for the continuation decision at the end of phase II, in the model above. The predicted values $\widehat{continuation_i} $ can be interpreted as the
probability of a drug moving to phase III conditional on the phase II z-score
(and other informative covariates observable at the end of phase II).

\subsection*{Kernel Density Estimation}

\label{supp4:kernel}

Let $Z_{1},Z_{2},\dots ,Z_{n}$ be the sample of z-score
in a given group of trials. To estimate the density we use the standard
weighted kernel estimator 
\begin{equation*}
\hat{f}(z)=\frac{1}{W}\sum_{i=1}^{n}\frac{w_{i}}{h}K\left( \frac{z-Z_{i}}{h}%
\right) ,
\end{equation*}%
where $W=\sum_{i=1}^{n}w_{i}$, $K(\cdot )$ is the Epanechnikov kernel
function, and $h$ the bandwidth which we choose with the Sheather-Jones
plug-in estimator \cite{Sheather1991}. To estimate the actual phase II
and phase III densities, we set all weights $w_{i}$ equal to one. To construct
the hypothetical densities controlled for \textit{selective continuation},
we estimate the kernel density of the phase II statistics, using the
predicted probabilities from our selection function as weights, i.e. $w_{i}=%
\widehat{continuation_{i}}$. The resulting densities for precisely reported
(i.e., not as inequality) test statistics by different groups of sponsors
are plotted in \hyperref[fig:densities]{Figure \ref*{fig:densities}}.

This procedure is similar in spirit to the weight function approach used to
test for publication bias in meta-analyses \cite{Hedges1992,Andrews2019}, but it
allows the weights to depend on more than one variable. The construction of
counterfactual distributions by weighted kernel density estimation has also
been used in other strands of the economics literature, e.g., for the
decomposition of the effects of institutional and labor market factors on
the distribution of wages \cite{DiNardo1996}.

\clearpage

\phantomsection
\addcontentsline{toc}{section}{References}

\printbibliography[notkeyword=SI]

\clearpage


\clearpage

\setcounter{table}{0}
\renewcommand{\thetable}{S\arabic{table}}
\setcounter{figure}{0}
\renewcommand{\thefigure}{S\arabic{figure}}

\section*{Supporting Information (SI)}

\subsection*{p-z Transformation}

\label{supp1:pz} Our analysis focuses on the reported p-values for the
statistical evaluation of trial results. However, the
p-density is not particularly well suited to perform discontinuity tests at the significance threshold because it is highly nonlinear in the relevant range. Neither is the p-density well suited for graphical representation, given that it is not possible to display both the region around the significant threshold and the overall distribution conveniently in the same graph.

To overcome these problems, we
transform the p-values to corresponding z-statistics by supposing that all
p-values would originate from a two-sided Z-test of a null hypothesis that
the drug has the same effect as the comparison. Given that under the null
hypothesis this statistic is normally distributed, we have the one-to-one
correspondence $z=-\Phi ^{-1}(\frac{p}{2})$, where $z$ is the absolute value
of the test-statistic and $\Phi ^{-1}$ is the inverse of the standard normal
cumulative distribution function. This transformation \textquotedblleft
stretches\textquotedblright\ the distribution from the $[0,1]$ interval to the
whole positive real axis with smaller p-values being stretched more. Hence, the region close to the significance threshold becomes more
prominent without losing the other parts of the distribution. Moreover, in the range around the significance threshold the z-density is close to linear, making it easier to identify discontinuities \cite{Cattaneo2019,Cattaneo2018}. A similar
transformation has been applied in the literature on experimental
biases across life sciences \cite{Holman2015}.

Note that the p-values in the dataset originate from diverse statistical
procedures (e.g., ANCOVA, ANOVA, Chi-squared-test, mixed models analysis,
linear regression, logistic regression, 1-sided t-test, 2-sided t-test,
etc.), with test statistics that follow different distributions, some
continuous, some discrete. Even though the sample size of the trials is
sufficiently large that, according to the Central Limit Theorem, many of the
resulting statistics are approximately normally distributed, in general the
actual test-statistic of the trial and our calculated z do not coincide.
Nevertheless, the p-z transformation allows us to conveniently compare the
results of all trials.

\begin{figure}[!p]
	\centering
	\includegraphics[width=\linewidth]{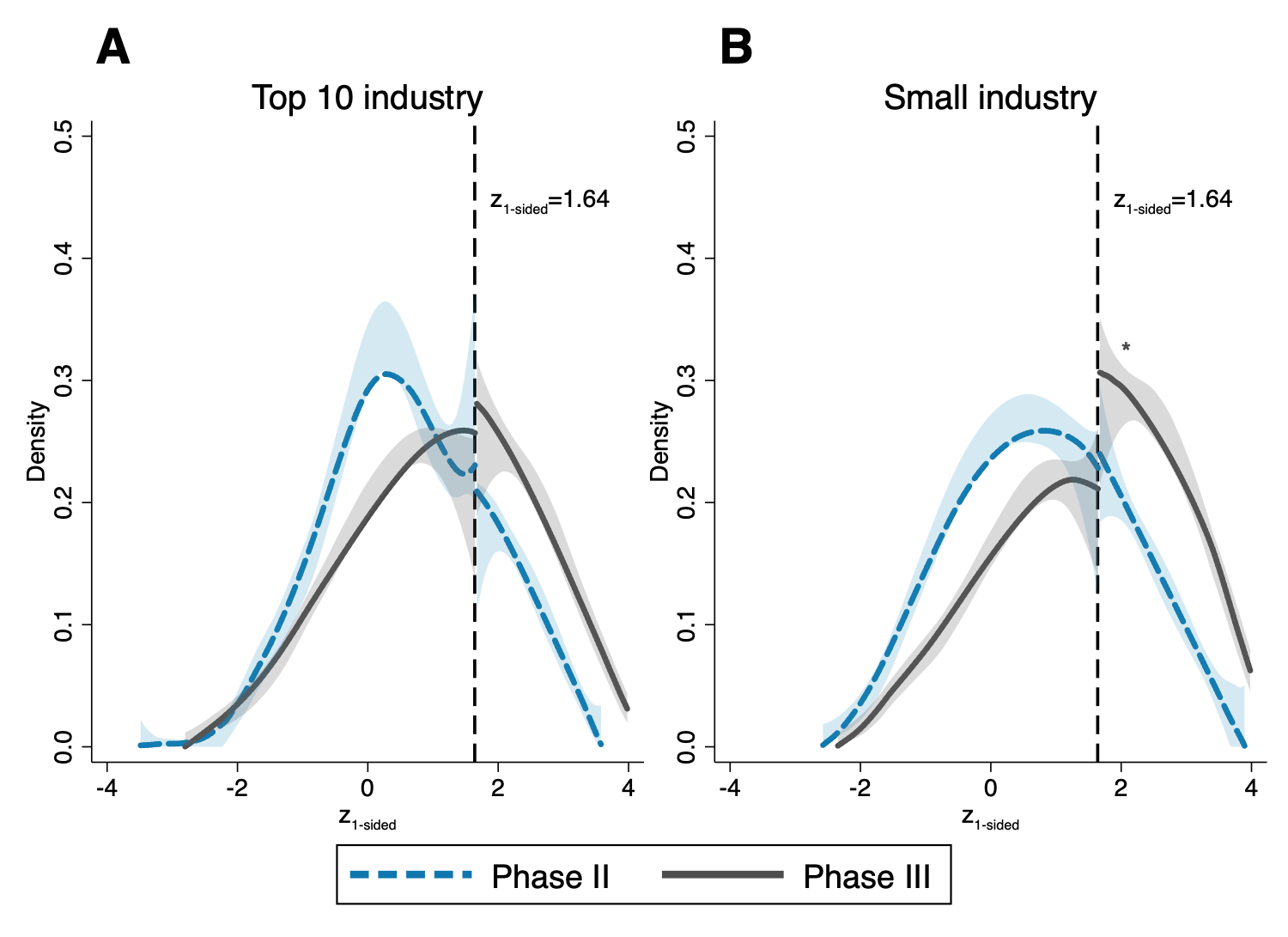}
	\caption{\textbf{Robustness check: density discontinuity tests for large and small industry sponsored trials with transformation to \emph{one-sided} test scores.} Density estimates of constructed one-sided z-statistics for primary
		outcomes of phase II (dashed blue lines) and phase III (solid grey lines)
		trials. The shaded areas are 95\%-confidence bands and the vertical lines at
		1.64 correspond to the threshold for statistical significance at 0.05 level.
		Sample sizes: A: $n=1,332$ (phase II), $n=1,424$ (phase III); B: $n=1,450$ (phase II), $n=1,520$ (phase III). Significance levels for discontinuity tests 
		\cite{Cattaneo2019}: * $p<0.1$; ** $p<0.05$; *** $p<0.01$; exact p-values reported in \hyperref[tab:cattaneo1s]{Table \ref*{tab:cattaneo1s}}.}
	\label{fig:discontinuities1s}
\end{figure}

To alleviate concerns that the discontinuity we find in the z-density for phase III trials by small industry sponsors (panel D of \hyperref[fig:discontinuities]{Figure \ref*{fig:discontinuities}} in the main text) may be driven by the specific transformation we choose, we provide density discontinuity tests for industry sponsored trials with p-values transformed to one-sided instead of two-sided test statistics. That is, $z_{1-sided}=-\Phi ^{-1}(p)$. The results, displayed in \hyperref[fig:discontinuities1s]{Figure \ref{fig:discontinuities1s}} and \hyperref[tab:cattaneo1s]{Table \ref*{tab:cattaneo1s}}, resemble closely those relying on the transformation to two-sided z-scores. We still find a sizable and statistically significant upward shift at the classical significance threshold for phase III trials by small sponsors. Also, the densities for phase III top ten and phase II (both types of sponsors) are smooth.

\subsection*{The Missing Tail of the z-Distribution}

\label{supp1:tail} Not all p-values in the registry are reported precisely,
but some are only stated in comparison to a certain threshold, e.g. $p<0.05$
or $p>0.1$. Whereas for most parts of the distribution this is a minor issue
and affects only a small number of observations, relative reporting becomes
the rule for very low p-values, corresponding to high z-statistics. In
particular, 30.8\% of the p-values in our sample of tests for primary
outcomes are reported as $p<0.001$ (corresponding to $z>3.29$) or $p<0.0001$
(corresponding to $z>3.89$). There are barely any p-values reported with
equality below these thresholds. For the z-distribution, this implies that
we know the size of the right tail (i.e., the mass above a certain
threshold) but we do not have any information about the exact shape.

For our analysis of the share of significant results, we deal with this
issue as following. As indicated in the regression equation, we include the
dummies $D1$ for ``$z>3.29$'' and $D2$ for ``$z>3.89$'' into the estimation
of the selection function, so that the probability of continuation is
estimated separately for those two cases. Moreover, we include p-values
which are reported as exactly zero (as a result of rounding) and hence
cannot be transformed into a z-score in the group $D2$. For the few cases in
which a z-score is reported as inequality with respect to a level $\bar{z}$
other than 3.29 and 3.89, we replace the respective $z$ with the mean of the
precisely reported z-statistics conditional on being above or respectively
below $\bar{z}$.

\begin{figure}[!p]
	\centering
	\includegraphics[width=.9\linewidth]{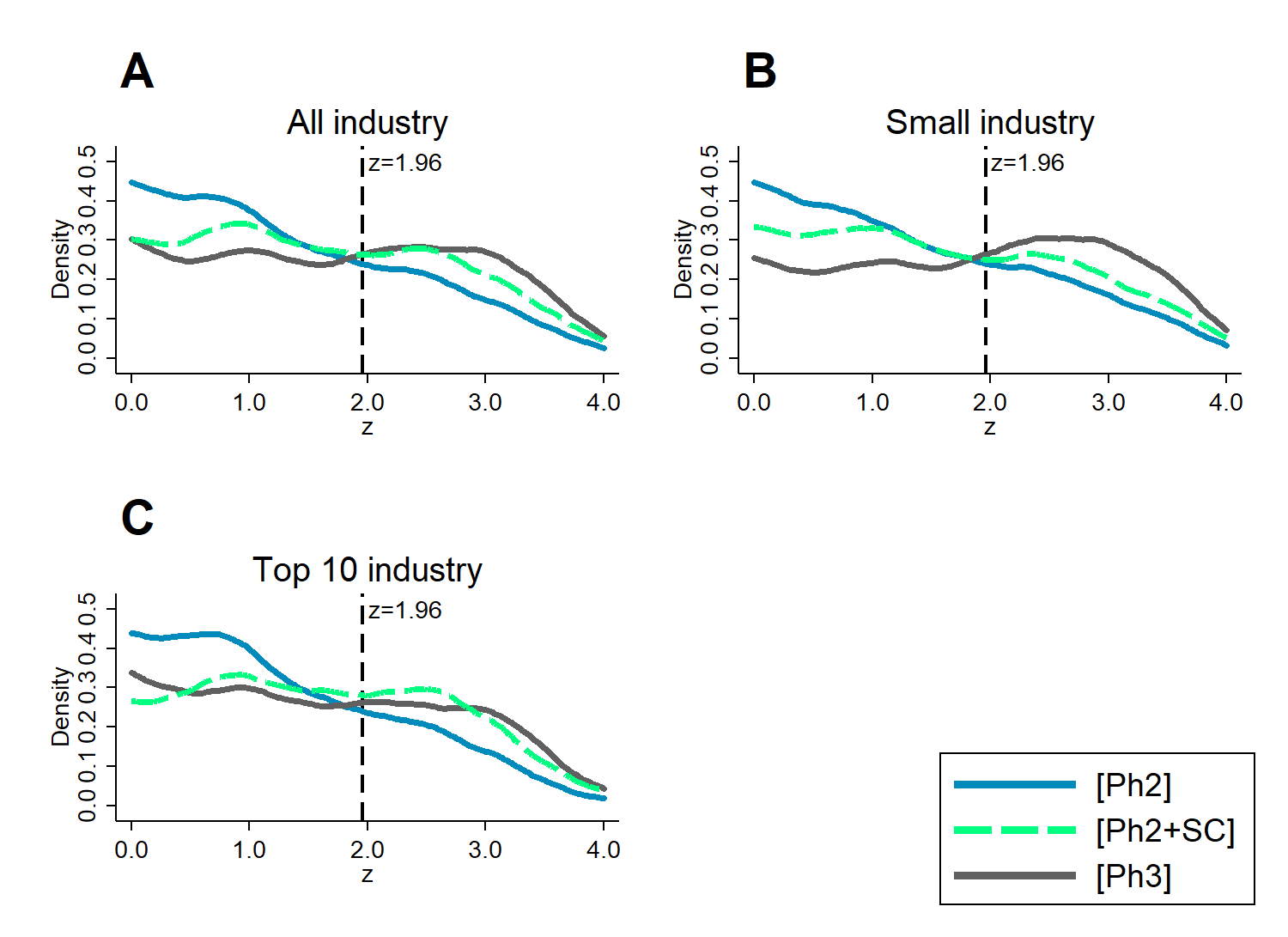}
	\caption{\textbf{Kernel density estimates for phase II and phase III z-scores
			and constructed counterfactuals accounting for \textit{selective
				continuation }, depending on affiliation of lead sponsor.} Estimated
		densities based only on p-values which are reported precisely (i.e. not as
		inequality). Shorthand notation: Ph2=phase II, Ph3=phase III, and SC=\textit{%
			selective continuation}. Sample sizes: A: $n=4,135$ (phase II), $n=5,957$ (phase III); B: $n=2,181$ (phase II), $n=3,209$ (phase III); C: $n=1,954$ (phase II), $n=2,748$ (phase III).}
	\label{fig:densities}
\end{figure}

For the discontinuity tests (\hyperref[fig:discontinuities]{Figure \ref*{fig:discontinuities}}, Figures \hyperref[fig:rob_disc]{\ref*{fig:rob_disc}}--\hyperref[fig:discontinuities_sec]{\ref*%
	{fig:discontinuities_sec}}, and Tables \hyperref[tab:cattaneo1]{ \ref*{tab:cattaneo1}}--\hyperref[tab:cattaneo1_sec]{\ref*{tab:cattaneo1_sec}}) and plots of densities (\hyperref[fig:densities]%
{Figure \ref*{fig:densities}}), we consider only p-values
which are reported precisely (i.e., not as inequality).

\subsection*{The Definition of Large vs. Small Industry Sponsors}
\label{def_large}
As our analysis relies on the estimation of densities, comparing trials by different groups of sponsors requires a discrete split of the sample. We focus on the impact of the size of the sponsoring corporations on their incentives. Therefore, we need a definition of ``large vs. small'' sponsors. In our main analysis, we compare the top ten sponsors in terms of 2018 revenues to the remaining smaller sponsors. These top ten are the ten companies in italics in the first column of \hyperref[tab:rankings]{Table \ref*{tab:rankings}}.
This particular definition is not only salient but also splits the sample of p-values roughly in half, maximizing statistical power in both subsamples. This is of particular importance for the density discontinuity tests, which require large sample sizes to be reliable.

To check robustness, we repeat our analysis for 56 alternative definitions of ``large'' and show that our main results hold across this wide range of alternatives for splitting the sample. As displayed in \hyperref[tab:rankings]{Table \ref*{tab:rankings}}, we rank sponsors not only by their 2018 revenues (column 1), but also by the volume of prescription drug sales in 2018 (column 2), R\&D spending in 2018, and the number of trials reported to the registry (column 4). It is not surprising that these four rankings are correlated.  For each of the four criteria, we create fourteen different definitions of ``large vs. small'': top seven vs. remainder, top eight vs. remainder, and so on up to top twenty vs. remainder. Hence, overall we have 14*4=56 different definitions, one of which is the top ten revenues definition we use for our main analysis.

\begin{figure}[!p]
	\centering
	\includegraphics[width=\linewidth]{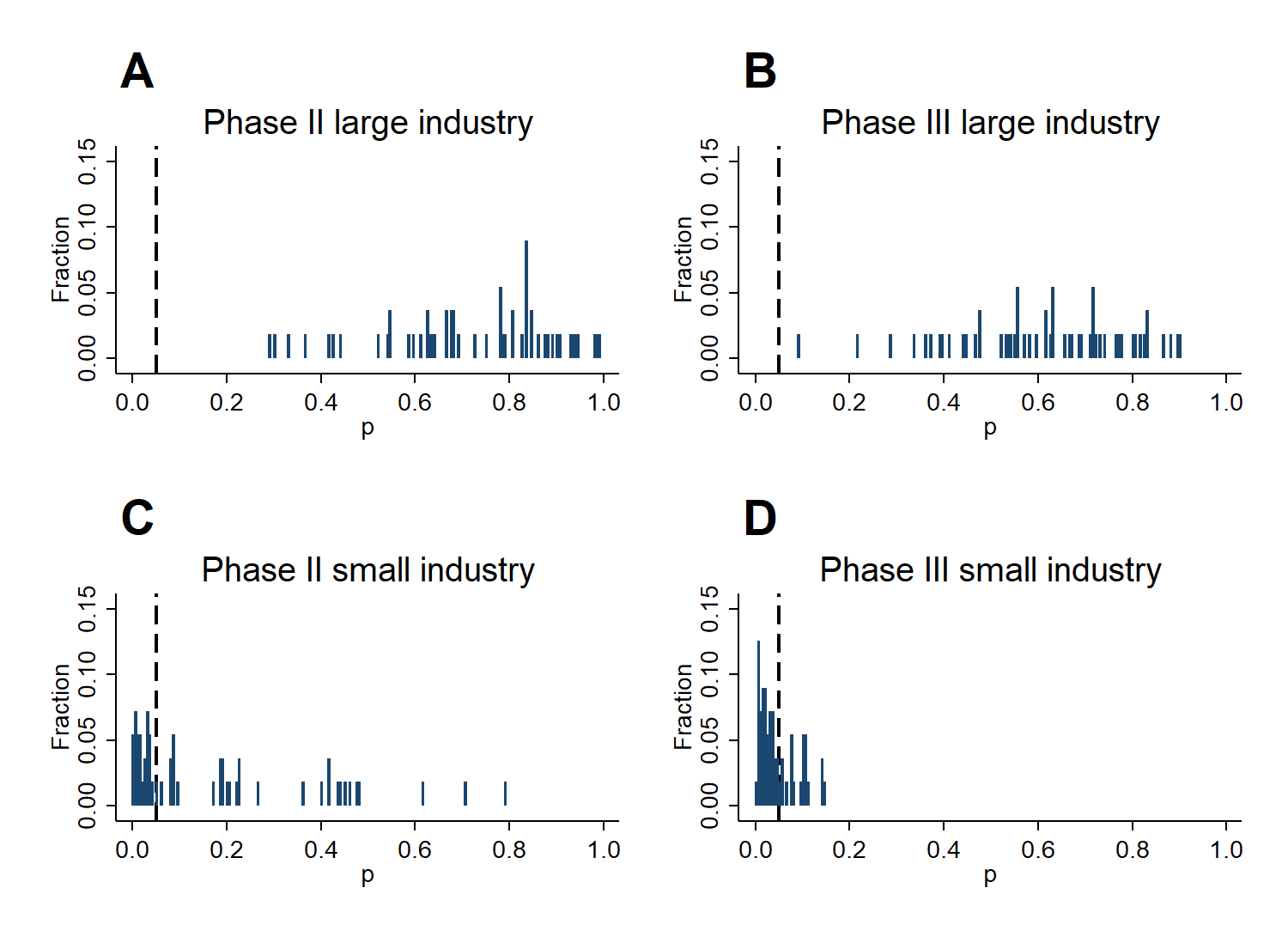}
	\caption{\textbf{Robustness check: histograms of p-values from density discontinuity tests at $z=1.96$ across 56 different definitions for large vs. small industry sponsors.} The p-values result from discontinuity tests \cite{Cattaneo2019} at $z=1.96$ in the densities of constructed z-statistics for primary outcomes. The dashed vertical lines indicate $p=0.05$. The sample of industry sponsored trials is split according to 56 different definitions of large sponsors. These definitions are obtained by ranking sponsors by their 2018 revenue, volume of prescription drug sales in 2018, R\&D spending in 2018, and the number of trials reported to the registry. For each of these four criteria, 14 different definitions of ``large vs. small'' are created: top seven vs. remainder, top eight vs. remainder, and so on up to top twenty vs. remainder. Further details are provided in the \hyperref[supp1:pz]{\textit{supplementary text}}. }
	\label{fig:rob_disc}
\end{figure}

\hyperref[fig:rob_disc]{Figure \ref{fig:rob_disc}} shows histograms of the p-values of density discontinuity tests across these 56 different definitions. In panels A and B we can see that the phase II and phase III z-densities for large industry sponsor never exhibit significant breaks at the 1.96 threshold, no matter which definition we use. As shown in panel C, for a number of definitions we find a significant discontinuity for phase II small industry, but at the same time in many cases we have p-values far above 0.05. Phase III small industry (panel D) is the only subgroup for which we find a significant break in our main specification. For the great majority of alternative definition, this finding is confirmed and the p-value never exceeds 0.146. 

We also repeat the counterfactual exercise of predicting the share of significant phase III results based on \textit{selective continuation} for each of the 56 different definitions. As discussed in the main text, the different patterns between large and small industry sponsors are robust across this wide range of alternative ways to define ``large'' sponsors (\hyperref[fig:decomposition]{Figure \ref{fig:decomposition}}, panels B and C).

\subsection*{Testing for Discontinuities of Distributions and Densities of z-scores}

\label{supp2} We provide a formal test of discontinuity in the z-score
density at the $z=1.96$ significance threshold. We implement manipulation
tests based on a state-of-the-art procedure developed by Cattaneo, Jansson,
and Ma \cite{Cattaneo2019,Cattaneo2018}. This test builds on a local polynomial
density estimation technique that avoids pre-binning of the data. \hyperref[tab:cattaneo1]%
{Table \ref*{tab:cattaneo1}} shows the p-values of the tests
performed on the densities from primary outcomes, depending on the
affiliation of the lead sponsors of the trials, as described in the main
text. We do not find any evidence of manipulation for trials in phase II. For
phase III, the p-values are lower, but when splitting the sample only significant for trials sponsored
by small industry.

\hyperref[fig:discontinuities]{Figure \ref*{fig:discontinuities}} in the
main text suggests that the breaks we find are not due to a spike, i.e., a
concentration of mass right above 1.96 (leading to a discontinuity in both
the density and the cumulative distribution function), but due to a
persistent upward shift in the density with an increased frequency of
results also further to the right of 1.96 (leading to a discontinuity only
in the density but not in the cumulative distribution function). To
reinforce this claim and distinguish the two cases, we perform further
density discontinuity tests with cutoffs 0.05 and 0.5 above the significance
threshold, corresponding to $z=2.01$ and $z=2.46$, for industry sponsored phase III
trials, for which we found a break at 1.96.

With this method we can implicitly test for a discontinuity in the
cumulative distribution function. If the discontinuity in the density was
due to a spike at 1.96, we would expect our test to find a downward jump in
the density at some point above. If there was manipulation and all inflated
results were concentrated exactly at 1.96 (sharp discontinuity in the
cumulative distribution function at 1.96), we should have a sharp downward
discontinuity in the density right above the threshold (captured by the test
at 2.01). Assuming more realistically that investigators want to push their
results above the significance threshold but cannot perfectly target a
p-value of 0.05, we would expect an excess mass above 1.96 that slowly
vanishes (captured by the test at 2.46). Even in the absence of a sharp
discontinuity, also in this case we would expect a downward tendency in the
density.

The differences of the bias-corrected density estimates to the right and to
the left of the respective cutoffs tabulated in \hyperref[tab:cattaneo2]{%
	Table \ref*{tab:cattaneo2}} do not display such a downward
tendency. To the contrary, for small industry sponsors, the differences at
2.01 and 2.46 have still a positive sign, the latter being even statistically significant. These findings confirm that there is a
persistent upward shift in the density around the significance threshold,
but there is no break in the cumulative distribution function with an excess
mass concentrated only just above 1.96.

\begin{figure}[!p]
	\centering
	\includegraphics[width=.9\linewidth]{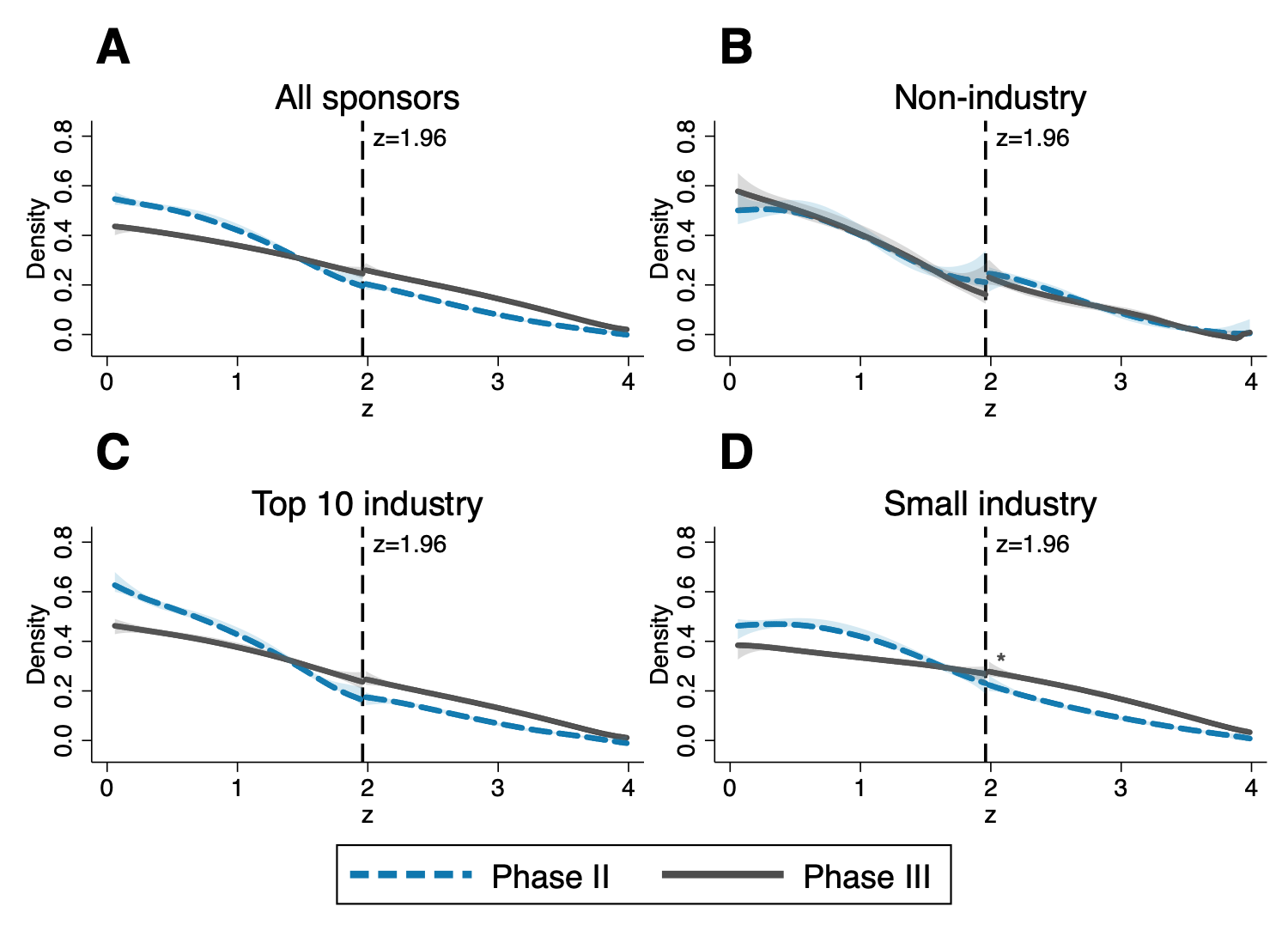}
	\caption{\textbf{Comparison of phase II and phase III z-score distributions and
			test for a discontinuity at $z=1.96$ for \emph{secondary outcomes}, depending on
			affiliation of lead sponsor.} Density estimates of the constructed
		z-statistics for tests on secondary outcomes of phase II (dashed blue lines)
		and phase III (solid grey lines) trials. The shaded areas are 95\%-confidence
		bands and the vertical lines at 1.96 correspond to the threshold for
		statistical significance at 0.05 level. Sample sizes: A: $n=17,840$ (phase II), $n=25,050$ (phase III); B: $n=2,553$ (phase II), $n=2,102$ (phase III); C: $n=8,579$ (phase II), $n=11,480$ (phase III); D: $n=6,672$ (phase II), $n=11,486$ (phase III).
		Significance levels for discontinuity tests \cite{Cattaneo2019}: * $%
		p<0.1 $; ** $p<0.05$; *** $p<0.01$; exact p-values reported in \hyperref[tab:cattaneo1_sec]{Table \ref*{tab:cattaneo1_sec}}.}
	\label{fig:discontinuities_sec}
\end{figure}

Similar discontinuity tests for the z-density from secondary outcomes do not
display any noteworthy break at the significance threshold (\hyperref[fig:discontinuities_sec]%
{Figure \ref*{fig:discontinuities_sec}} and \hyperref[tab:cattaneo1_sec]%
{Table \ref*{tab:cattaneo1_sec}}). Moreover, the excess mass
of significant results from industry-sponsored trials in phase III relative to
phase II is much smaller compared to the distribution for primary outcomes.

\subsection*{Linking Phase II and Phase III Trials}

\label{supp3}

To analyze \textit{selective continuation} from phase II to
phase III, we link phase II and phase III trials in our dataset, based on the main intervention, the medical condition to be treated, and the timing. This is not such a straightforward exercise to
implement for two reasons:

\begin{itemize}
	\item The AACT dataset is a mere digitization of the reported trial protocols.
	Hence, most variables are not well codified and have non-generic entries.
	Even though the information on interventions and conditions of the
	trials for which results are reported is rather complete, the cells in the
	reporting forms are interpreted differently by different reporting parties.
	For instance, in the specification of a trial's intervention, in many cases
	all the drugs involved in the trial are inserted in one cell, without
	specifying whether the drugs are given as a combination or separately to
	different arms of the trial. Often, it is not specified which drug
	constitutes the experimental treatment rather than the control. Hence, it is
	not possible to mechanically identify a trial's main experimental
	intervention. As an additional complication, many drugs appear in the data
	with different names; some times the drugs are referred by the chemical
	composition, while other times by their commercial name.
	
	\item The process of drug development is not linear in the sense that we
	usually do not have one phase II trial followed by one phase III trial and then
	a request for FDA approval. In most cases, there is a number of phase II
	trials looking at similar but potentially slightly different
	interventions/conditions, such as different drug dosages, different
	characteristics of eligible patients, or different control interventions.
	These phase II trials are typically followed by an even larger number of phase III trials
	with similar interventions/conditions but slightly varying specifications.
\end{itemize}

We address these hurdles in the following way. We read one by one the
protocols for all the phase II trials in the dataset for which at least one p-value is
reported and which were completed before end of December 2018. With this restriction on the
completion date, there could potentially be a follow-up phase III trial registered before
August 2019. From the protocols, we determine the main
experimental intervention(s), i.e., the main drug or combination of drugs
whose efficacy and safety is to be established, for 1,773 phase II trials. As indication of the medical condition the trials address, we use the Medical Subject Headings (MeSH) terms determined by the curators for the purpose of
making the \textit{ClinicalTrials.gov} webpage searchable \cite{Tasneem2012}, disregarding overly generic categories such as
simply \textquotedblleft Disease\textquotedblright.

We consider a phase II trial as continued if we could link it to at least one phase III trial. That is, if we found at least one phase III trial registered in the
database (regardless of whether associated results are reported or not)
fulfilling all of the following criteria:

\begin{enumerate}
	\item \textbf{Intervention:} All drugs being part of at least one of the determined main
	interventions of the phase II trial appear as listed interventions in the
	phase III trial. This is either with exactly the same name or with a synonym
	which the reporting party states to refer to the same drug.
	
	\item\textbf{Condition:} All the MeSH-conditions associated to the phase II trial are also
	associated to the phase III trial.
	
	\item \textbf{Timing:} The start date of the phase II trial was before the start date of the
	phase III trial.
\end{enumerate}

This linking is not perfect, for instance because it disregards whether all
the drugs in the phase III trial were part of one combination in one arm.
Moreover, we do not take into consideration other details of the trials like the exact population of eligible patients.
However, given the limitations of the data, this procedure appears
reasonably accurate. We manage to link 33.3\% of the industry-sponsored
phase II trials in our restricted dataset to at least one phase III trial.
These numbers are in line with the ones reported in previous studies \cite{DiMasi2003} and on the FDA webpage \cite{FDA2018}. For
non-industry sponsored trials, however, reporting in phase III is very meager
and we can find phase III matches for only 18.0\% of the phase II trials. Given
this low number and the fact that there are no significant differences
between the phase II and phase III distribution for non-industry sponsors to
begin with, we investigate selection only for industry-sponsored trials.

Note that criterion 3 considers only the start dates of the trials. It might
appear to be more intuitive to require the completion date of the phase II
trial to be prior to the start date of the phase III trial. Indeed, most of
our linked trials fulfill also this stronger condition. However, in some
cases this condition is too strong. That is, some phase III trials start
before the corresponding phase II trials are fully completed. For instance,
some phase II results on long-run impacts might still be pending but the
collected evidence is already strong enough for the investigators to start a
phase III trial. Moreover, we consider the reported start dates to be more
reliable. The responsible parties might have incentives to report a later
completion date than the actual, in order to meet the requirements for
timely reporting of results.

\subsection*{MeSH Condition Fixed Effects and Market Size Data}
\label{supp1:mkt}
To account for potential systematic differences across drugs for the treatment of different kinds of conditions, we include condition fixed effects in the estimation of the selection functions for \emph{selective continuation}. For this purpose, we assign each trial in one of the 15 largest categories of conditions (in terms of frequency in our data), based on the MeSH terms determined by the curators of the database \cite{Tasneem2012}. These categories are displayed in \hyperref[tab:mktsize]{Table \ref*{tab:mktsize}}. Some strongly overlapping categories have been merged.
Trials that could not be assigned to a specific group or belong to one of the smaller groups constitute the omitted category. In case a trial is associated with more than one category, we assign it to the one with the largest expected market size.

To obtain a proxy for the expected market size for a newly
developed drug, we evaluate the Medicare D spending for existing drugs in
2011 according to information from the \textit{Centers for Medicare \&
	Medicaid Services} publicly available at \href{https://www.cms.gov/Research-Statistics-Data-and-Systems/Research-Statistics-Data-and-Systems.html}%
{%
	https://www.cms.gov/Research-Statistics-Data-and-Systems/Research-Statistics-Data-and-Systems.html%
}. \textit{Part D Prescription Drug Event} (PDE) data is provided for a
subset ($\sim 70\%$) of Medicare beneficiaries.

We classify manually 1,056 marketed drugs, among which the 420 with the
highest Medicare D spending, into the MeSH categories for the treated
conditions. Overall,
these drugs make up for 90\% of the expenditure on the drugs in the dataset. 
\hyperref[tab:mktsize]{Table \ref*{tab:mktsize}} shows also the
total spending by category.

\subsection*{Background on \textit{ClinicalTrials.gov}}

\label{supp1:data} \textit{ClinicalTrials.gov} is an online registry of
clinical research studies in human volunteers. The website is maintained by
the \textit{National Library of Medicine} (NLM) at the \textit{National
	Institutes of Health} (NIH) in collaboration with the \textit{U.S. Food and
	Drug Administration} (FDA). It was established in February 2000 with the aim
to increase transparency in clinical research. Initially, the registry
contained only trials to test the efficacy of new experimental drugs for
serious or life-threatening diseases or conditions and registration was
mainly voluntary. For more information on the history of the registry,
related policies, and laws, see \href{https://clinicaltrials.gov/ct2/about-site/history}%
{https://clinicaltrials.gov/ct2/about-site/history} (accessed Jun 23, 2017).

In 2007 the requirements for registration of trials were extended
substantially through the \textit{FDA Amendments Act} (FDAAA) \cite{Wood2009}. Even though in January 2017 those rules have been
redefined more precisely \cite{Zarin2016}, in the following we will
refer to the regulation of Section 801 of the FDAAA which was the
legislation in force at the time when the great majority of the data in our
analysis was generated. According to \href{https://clinicaltrials.gov/ct2/manage-recs/fdaaa}{
	https://clinicaltrials.gov/ct2/manage-recs/fdaaa} (accessed Jun 23, 2017), the main criteria a trial must meet to be affected
by this regulation are the following:

\begin{itemize}
	\item initiated after September 27, 2007, or initiated on or before that
	date and still ongoing as of December 26, 2007;
	
	\item controlled clinical investigation of drugs, biologics or medical
	devices other than phase I trials and small feasibility studies;
	
	\item the trial has one or more sites in the United States or it involves
	drugs, biologics or medical devices manufactured in the United States.
\end{itemize}

If these criteria apply, the responsible party (i.e., the sponsor or the
principal investigator of the trial) must register the trial and provide the
required information no later than 21 days after enrollment of the first
participant. In case the investigated drug, biologic, or device is approved,
licensed, or cleared by FDA, moreover, the responsible party must submit
some basic summary results of the trial no later than twelve months after
the completion date. Since September 2008, these submitted results are publicly accessible in the \textit{ClinicalTrials.gov} results database so as to reach an even higher level of transparency.
However, there are some loopholes in the legislation \cite{Zarin2008}; for instance, the required level of details of the results is not clearly defined and phase I trials and trials of not-approved products are exempt. In all the other
cases that do not meet the stated criteria, registering and reporting of
results is voluntary.

The FDAAA establishes penalties for non-compliance of up to \$10,000 per
day. However, no enforcement has yet occurred \cite{Anderson2015,Piller2018,DeVito2020,Piller2020}. Assessing compliance rates is not easy because the
aforementioned exemptions and imprecisions in the FDAAA legislation
complicate identifying which trials are applicable. An early algorithm-based
study \cite{Anderson2015} shows that only 13.4\% of applicable clinical
trials registered on \textit{ClinicalTrials.gov} between 2008 and 2012
reported results in a timely fashion and only 38.3\% reported results at any
time at all. However, in a manual review of a subsample of trials the same
authors \cite{Anderson2015} found that their methodology based on
assumptions about the approval status of the drug tended to underestimate
reporting rates. Later studies document for a sample of 329
industry-sponsored phase II-IV US trials completed or terminated 2007-2009 a
result reporting rate to \textit{ClinicalTrials.gov} of 58\% by December 2014 \cite{Zarin2017b} and an increase of the overall reporting rate for
applicable trials from 58\% to 72\% in the two years before September
2017, driven not by fear of sanctions but by public pressure on the
responsible parties \cite{Piller2018}.

Since January 2017 the improved \textquotedblleft Final
Rule\textquotedblright\ is in place (hence, it does not affect the great
majority of the trials we analyze), addressing many loopholes and broadening
the scope of the 2007 legislation \cite{Zarin2016}. However, the FDA's
efforts to police compliance are still very limited \cite{Piller2018,DeVito2020,Piller2020}.
Beyond the disclosure mandate, the FDAAA raised public awareness about the
importance of transparency in clinical research and led many large
pharmaceutical companies and research institutions to develop internal
disclosure policies \cite{Anderson2015,
	Piller2018,Piller2020}.

The most recent and complete evaluation of compliance with the FDAAA Final Rule finds 64.5\% of industry-sponsored trials to report any results and 50.3\% to be fully compliant with the rules; that is, to report results within one year of the primary completion date \cite{DeVito2020}.

Considering the missing enforcement of the FDAAA regulations, lack of
reporting does not necessarily mean that the responsible parties intend to
hide their results, but rather that they just do not take the time to go
through the lengthy reporting process. In this light, notwithstanding the
legal requirements, for the purpose of our analysis reporting of results
should be seen as mostly voluntary.

Several studies in the medical literature assess the quality of the data
reported to the registry and the results database along different dimension,
e.g., information about scientific leadership \cite{Sekeres2008},
consistency of reported primary outcomes \cite{Mathieu2009,
	Ramagopalan2014}, comparisons to results published in academic journals \cite{Zarin2017b, Ross2009}, and the provision of Individual
Participant Data (IPD) \cite{Zarin2016b}. All these studies, as well as
overall assessments by the curators of the database \cite{Zarin2017,
	Zarin2011}, find ambiguous results and see scope for improvement \cite{Dickersin2018}.

The biggest challenge when working with the AACT data is that, as a mere
digitization of the trial protocols, most variables have non-generic entries
and many of them contain large bodies of text. Moreover, reporting parties do not always
interpret the different cells in the reporting form in the same way. For
instance, when reporting the intervention of a trial, in many
cases all the drugs involved in a trial are inserted in one cell without
specifying whether they are given as a combination or separately to
different arms of the trial. Furthermore, reporting parties
indicate differently which drug constitutes the experimental treatment and which one is the
control. Often, one can find a clarification in other parts
of the protocol. Similar issues arise with many of the self-reported
variables. Even though for most trials the reported content is complete and the whole study protocol embedded in the context gives a clear
picture, different parties often report the same information in different cells. This non-uniformity prevents the mechanical evaluation of large parts of the data, even with natural language processing algorithms.

Consequently, we are forced to either codify the data by hand (like the main
intervention of phase II trials which we use for our linking of trials across
phases) or restrict attention to characteristics that are codified uniformly
among all the trials in the database. The latter are numerical
entries or entries that allow only for a finite, prespecified number of
answers (e.g., binary variables).

We classify trials and link them across phases based on the MeSH terms associated to the treated conditions. The MeSH thesaurus is a controlled list of
vocabulary produced by the \textit{National Library of Medicine} and used
for indexing, cataloging, and searching biomedical and health-related
information. The MeSH classification is provided by \textit{ClinicalTrials.gov} administrators based on natural language processing algorithms.

\clearpage

\clearpage

\begin{sidewaystable}[!p]
	\caption{\textbf{Ranking of Industry Sponsors by Different Criteria.}}
	\label{tab:rankings}
	\centering
	\begin{tabular}{ccccc}
		\toprule
		\textbf{Rank} & \textbf{Revenue 2018} & \textbf{Rx Sales 2018} & \textbf{R\&D Spending 2018} & \textbf{No. Trials Reported} \\
		\midrule
		1     & \textit{Johnson \& Johnson} & Pfizer & Roche & GlaxoSmithKline \\
		2     & \textit{Roche} & Roche & Johnson \& Johnson & Pfizer \\
		3     & \textit{AbbVie/Abbott Laboratories} & Novartis & Novartis & Merck Sharp \& Dohme Corp. \\
		4     & \textit{Pfizer} & Johnson \& Johnson & Pfizer & Eli Lilly \& Co \\
		5     & \textit{Novartis} & Merck Sharp \& Dohme Corp. & Merck Sharp \& Dohme Corp. & Boehringer Ingelheim \\
		6     & \textit{Bayer} & AbbVie/Abbott Laboratories & Sanofi & AstraZeneca \\
		7     & \textit{GlaxoSmithKline} & Sanofi & AbbVie/Abbott Laboratories & Roche \\
		8     & \textit{Merck Sharp \& Dohme Corp.} & GlaxoSmithKline & AstraZeneca & Novartis \\
		9     & \textit{Sanofi} & Amgen & Bristol-Myers Squibb & Takeda Pharmaceutical \\
		10    & \textit{Eli Lilly \& Co} & Gilead Sciences & Eli Lilly \& Co & Shire \\
		11    & Amgen & Bristol-Myers Squibb & GlaxoSmithKline & Amgen \\
		12    & Bristol-Myers Squibb & AstraZeneca & Celegne & Bayer \\
		13    & Gilead Sciences & Eli Lilly \& Co & Gilead Sciences & Sanofi \\
		14    & AstraZeneca & Bayer & Amgen & Johnson \& Johnson \\
		15    & Danaher Corporation & Novo Nordisk & Bayer & Gilead Sciences \\
		16    & Boehringer Ingelheim & Takeda Pharmaceutical & Boehringer Ingelheim & Bristol-Myers Squibb \\
		17    & Takeda Pharmaceutical & Celegne & Takeda Pharmaceutical & Otsuka Holdings \\
		18    & Teva Pharmaceutical Industries & Shire & Biogen & AbbVie/Abbott Laboratories \\
		19    & Novo Nordisk & Boehringer Ingelheim & Novo Nordisk & Novo Nordisk \\
		20    & Merck KGaA & Allergan & Regeneron Pharmaceuticals & Merck KGaA \\
		\bottomrule
	\end{tabular}
	\caption*{\small \textit{Notes:} The companies in italics are defined as the top ten industry sponsors in our main analaysis. Known large-scle subsidiaries are grouped with their mother corporation (e.g. Janssen Research \& Development as part of Johnson \& Johnson). Small companies that may have alliances with (or are later acquired by) larger companies are coded as separate sponsors. Shire was acquired by Takeda Pharmaceutical in early 2019 but we treat the two companies separately, as this acquisitaion happend at the very end of our sample period. Sources: Revenue 2018: \href{https://en.wikipedia.org/wiki/List_of_largest_biomedical_companies_by_revenue}{https://en.wikipedia.org/wiki/List\_of\_largest\_biomedical\_companies\_by\_revenue} (revenue data collected from financial statements on company websites, accessed Oct 23, 2019); Rx (i.e. prescription drugs) Sales 2018 and R\&D Spending 2018: \cite{Christel2019} (based on data from EvaluatePharma\textregistered); No. Trials Reported: own calculations based on \textit{ClinicalTrials.gov} data.}
\end{sidewaystable}

\clearpage

\begin{table}[!p]
	\caption{\textbf{P-values for tests of density discontinuity at the $z=1.96$
			threshold.}}
	\label{tab:cattaneo1}
	\centering
	\begin{tabular}{lcc}
		\toprule
		& \textbf{(1)} & \textbf{(2)} \\
		\textbf{Sponsor} & \textbf{Phase II} & \textbf{Phase III} \\
		\midrule
		All   & 0.09*  & 0.00*** \\
		& (3,953) & (3,664) \\
		Non-industry & 0.23  & 0.35 \\
		& (1,171) & (720) \\
		All industry & 0.30   & 0.52 \\
		& (2,782) & (2,944) \\
		Small industry & 0.20   & 0.032** \\
		& (1,450) & (1,520) \\
		Top 10 industry & 0.91  & 0.67 \\
		& (1,332) & (1,424) \\
		\bottomrule
	\end{tabular}%
	\caption*{\small \textit{Notes:} P-values result from the density discontinuity test \cite{Cattaneo2019}, described in detail in the \hyperref[supp2]{\textit{supplementary text}}, for primary outcomes; significance levels: * $p<0.1$; ** $p<0.05$; *** $p<0.01$
		. Sample sizes in parentheses.}
\end{table}


\begin{table}[!p]
	\caption{\textbf{Size of discontinuities in the z-density at the
			significance threshold, as well as at $z=2.01$ and $z=2.46$, for
			industry-sponsored phase III trials.}}
	\label{tab:cattaneo2}
	\centering
	\begin{tabular}{lccc}
		\toprule
		\multicolumn{1}{r}{} & \multicolumn{1}{c}{\textbf{(1)}} & \multicolumn{1}{c}{%
			\textbf{(2)}} & \multicolumn{1}{c}{\textbf{(3)}} \\ 
		\textbf{Sponsor \textbackslash Cutoff value} & \multicolumn{1}{c}{\textbf{%
				z=1.96}} & \multicolumn{1}{c}{\textbf{z=2.01}} & \multicolumn{1}{c}{\textbf{%
				z=2.46}} \\ 
		\midrule
		All industry & 0.031 & 0.087** & 0.11* \\
		Small industry & 0.166** & 0.056 & 0.177** \\
		Top 10 industry & 0.029 & 0.075 & 0.015 \\
		\bottomrule 
	\end{tabular}%
	\caption*{\small \textit{Notes:} Differences of the bias-corrected density estimates to the
		right and to the left of the respective cutoff, resulting from the density
		discontinuity test \cite{Cattaneo2019}, described in detail in the \hyperref[supp2]{\textit{supplementary text}}, for primary outcomes. Sample sizes: All industry $%
		n=2,944$, Small industry $n=1,520$, Top 10 industry $n=1,424$. Significance
		levels: * $p<0.1$; ** $p<0.05$; *** $p<0.01$.}
\end{table}

\clearpage

\begin{table}[!p]
	\caption{\textbf{Robustness check -- transformation to \emph{one-sided} test scores: P-values for tests of density discontinuity at the $z_{1-sided}=1.64$ threshold.}}
	\label{tab:cattaneo1s}
	\centering
	\begin{tabular}{lcc}
		\toprule
		& \textbf{(1)} & \textbf{(2)} \\
		\textbf{Sponsor} & \textbf{Phase II} & \textbf{Phase III} \\
		\midrule
		Small industry & 0.20   & 0.076* \\
		& (1,450) & (1,520) \\
		Top 10 industry & 0.31  & 0.74 \\
		& (1,332) & (1,424) \\
		\bottomrule
	\end{tabular}%
	\caption*{\small \textit{Notes:} P-values result from the density discontinuity test \cite{Cattaneo2019}, described in detail in the \hyperref[supp2]{\textit{supplementary text}}, for primary outcomes; significance levels: * $p<0.1$; ** $p<0.05$; *** $p<0.01$
		. Sample sizes in parentheses.}
\end{table}

\begin{table}[!p]
	\caption{\textbf{P-values for tests of density discontinuity at the z=1.96
			threshold -- \emph{secondary outcomes}.}}
	\label{tab:cattaneo1_sec}
	\centering
	\begin{tabular}{lcc}
		\toprule
		& \textbf{(1)} & \textbf{(2)}\\
		\textbf{Sponsor} & \multicolumn{1}{c}{\textbf{Phase II}} & 
		\multicolumn{1}{c}{\textbf{Phase III}} \\
		\midrule 
		All   & 0.54  & 0.21 \\
		& (17,804) & (25,050) \\
		Non-industry & 0.34  & 0.35 \\
		& (2,553) & (2,102) \\
		All industry & 0.34  & 0.07* \\
		& (15,251) & (22,948) \\
		Small industry & 0.87  & 0.06* \\
		& (6,672) & (11,468) \\
		Top 10 industry & 0.44  & 0.36 \\
		& (8,579) & (11,480) \\
		\bottomrule
	\end{tabular}
	\caption*{\small \textit{Notes:} P-values result from the density discontinuity test \cite{Cattaneo2019}, described in detail in the \hyperref[supp2]{\textit{supplementary text}}, for secondary outcomes; significance levels: * $p<0.1$; ** $p<0.05$; *** $%
		p<0.01 $. Sample sizes in parentheses.}
\end{table}

\clearpage

\begin{table}[!p]
	\caption{\textbf{Estimates of logit selection function for \textit{selective
				continuation}, based on \emph{secondary outcomes}.}}
	\label{tab:selection_sec}
	\centering
	\begin{tabular}{lccc}
		\toprule
		& \textbf{(1)} & \textbf{(2)} & \textbf{(3)} \\
		\textbf{Sponsor} & \textbf{All} & \textbf{Small} & \textbf{Top 10} \\
		& \textbf{industry} & \textbf{industry} & \textbf{industry} \\
		\midrule
		&       &       &  \\
		Phase II z-score & 0.109* & 0.197** & -0.0612 \\
		& (0.0557) & (0.0839) & (0.0674) \\
		Dummy for phase II z-score reported as ``$z>3.29$'' & 0.465 & 0.600 & 0.0737 \\
		& (0.416) & (0.628) & (0.461) \\
		Dummy for phase II z-score reported as ``$z>3.89$'' & 0.512 & 0.279 & 0.779** \\
		& (0.353) & (0.395) & (0.351) \\
		Mean dependent variable & 0.353 & 0.360 & 0.347 \\
		&       &       &  \\
		Controls & yes   & yes   & yes \\
		MeSH condition fixed effects & yes   & yes   & yes \\
		Completion year fixed effects & yes   & yes   & yes \\
		Observations & 17,724 & 7,502 & 10,222 \\
		No. of trials & 720   & 402   & 318 \\
		\bottomrule
	\end{tabular}
	\caption*{\small \textit{Notes:} Unit of observation: trial-outcome;
		included controls: square root of the overall enrollment and dummy for placebo
		comparator. Categories for condition fixed effects are based on Medical Subject Headings (MeSH) terms associated to the trials \cite{Tasneem2012}; for more details, see \hyperref[supp1:mkt]{\textit{supplementary text}}. Standard errors in parentheses are
		clustered at the MeSH condition level; significance levels (based on a two-sided
		t-test): * $p<0.1$; ** $p<0.05$; *** $p<0.01$. }
\end{table}

\clearpage

\begin{table}[!p]
	\caption{\textbf{Selection-based decomposition of the difference in
			significant results from primary outcomes between phase II and phase III,
			depending on affiliation of lead sponsor.}}
	\label{tab:decomposition}
	\centering
	\small
	\begin{tabular}{lccc}
		\toprule
		\multicolumn{4}{c}{\textbf{Share of significant results}} \\ 
		\midrule
		& \textbf{(1)} & \textbf{(2)} & \textbf{(3)} \\ 
		\textbf{Sponsor} & \textbf{All} & \textbf{Small} & \textbf{Top 10} \\ 
		& \textbf{industry} & \textbf{industry} & \textbf{industry} \\ 
		\midrule
		&  &  &  \\ 
		{[}Ph2] & 0.481 & 0.499 & 0.460 \\
		& (0.0205) & (0.0226) & (0.0319) \\
		{[}Ph3] & 0.721 & 0.757 & 0.679 \\
		& (0.0149) & (0.0136) & (0.0245) \\
		{[}Ph2+SC] & 0.604 & 0.573 & 0.645 \\
		& (0.0316) & (0.0406) & (0.0458) \\ 
		\bottomrule
		&  &  &  \\ 
		\toprule
		\multicolumn{4}{c}{\textbf{Differences}} \\
		\midrule 
		& \textbf{(4)} & \textbf{(5)} & \textbf{(6)} \\ 
		\textbf{Sponsor} & \textbf{All} & \textbf{Small} & \textbf{Top 10} \\ 
		& \textbf{industry} & \textbf{industry} & \textbf{industry} \\ 
		\midrule
		&  &  &  \\ 
		{[}Ph3]-[Ph2] & 0.241*** & 0.258*** & 0.219*** \\
		& (0.0259) & (0.0256) & (0.0422) \\
		{[}Ph3]-[Ph2+SC] & 0.117*** & 0.184*** & 0.0339 \\
		& (0.0354) & (0.0426) & (0.0529) \\
		{[}Ph2+SC]-[Ph2] & 0.123*** & 0.0746** & 0.185*** \\
		& (0.0260) & (0.0347) & (0.0410) \\
		&       &       &  \\
		Observations & 10,092 & 5,390 & 4,702 \\
		Observations Ph2 & 4,135  & 2,181  & 1,954 \\
		Observations Ph3 & 5,957  & 3,209  & 2,748 \\
		No. of trials Ph2 & 1,244  & 732   & 512 \\
		No. of trials Ph3 & 2,655  & 1,544  & 1,111 \\ 
		\bottomrule
	\end{tabular}
	\caption*{\small \textit{Notes:} Columns 1-3 display the share of significant results based
		on kernel density estimates and adjustment for selection, with shorthand
		notation Ph2=phase II, Ph3=phase III, and SC=\textit{selective continuation}.
		Columns 4-6 display the differences in these shares. The standard errors in
		parentheses are obtained by bootstrapping the whole estimation procedure
		(500 repetitions, clustered at the trial level); significance levels (based on a two-sided t-test): * $
		p<0.1$; ** $p<0.05$; *** $p<0.01$. }
\end{table}

\begin{table}[!p]
	\caption{\textbf{Categories for MeSH condition fixed effects with market size determined from total Medicare D
			spending.}}
	\label{tab:mktsize}
	\centering
	\begin{tabular}{rrc}
		\toprule
		\multicolumn{1}{c}{\textbf{MeSH}} &  & \textbf{Total Medicare D} \\ 
		\multicolumn{1}{c}{\textbf{code}} & \multicolumn{1}{c}{\textbf{Category}} & 
		\textbf{Spending in 2011} \\ 
		&  & in bn US\$ \\ 
		\midrule
		&  &  \\ 
		\multicolumn{1}{l}{C14} & \multicolumn{1}{l}{Cardiovascular Diseases} & 
		13.215 \\ 
		\multicolumn{1}{l}{F03} & \multicolumn{1}{l}{Mental Disorders} & 12.336 \\ 
		\multicolumn{1}{l}{C18} & \multicolumn{1}{l}{Nutritional and Metabolic
			Diseases} & 8.957 \\ 
		\multicolumn{1}{l}{C19} & \multicolumn{1}{l}{Endocrine System Diseases} & 
		8.45 \\ 
		\multicolumn{1}{l}{C10} & \multicolumn{1}{l}{Nervous System Diseases} & 5.956
		\\ 
		\multicolumn{1}{l}{C08/C09} & \multicolumn{1}{l}{Respiratory Tract
			Diseases/Otorhinolaryngologic Disease} & 5.945 \\ 
		\multicolumn{1}{l}{C06} & \multicolumn{1}{l}{Digestive System Diseases} & 
		4.377 \\ 
		\multicolumn{1}{l}{C05} & \multicolumn{1}{l}{Musculoskeletal Diseases} & 
		2.888 \\ 
		\multicolumn{1}{l}{C04} & \multicolumn{1}{l}{Neoplasms} & 2.64 \\ 
		\multicolumn{1}{l}{C12/C13} & \multicolumn{1}{l}{Male Urogenital Diseases/}
		& 2.262 \\ 
		& \multicolumn{1}{c}{Female Urogenital Diseases and Pregnancy Complications}
		&  \\ 
		\multicolumn{1}{l}{C20} & \multicolumn{1}{l}{Immune System Diseases} & 1.355
		\\ 
		\multicolumn{1}{l}{C23} & \multicolumn{1}{l}{Pathological Conditions, Signs
			and Symptoms} & 0.812 \\ 
		\multicolumn{1}{l}{C17} & \multicolumn{1}{l}{Skin and Connective Tissue
			Diseases} & 0.683 \\ 
		\multicolumn{1}{l}{C25} & \multicolumn{1}{l}{Chemically-Induced Disorders} & 
		0.17 \\ 
		\multicolumn{1}{l}{C16} & \multicolumn{1}{l}{Congenital, Hereditary, and
			Neonatal Diseases and Abnormalities} & 0.101 \\ 
		&  &  \\ 
		\bottomrule 
	\end{tabular}
	\caption*{\small \textit{Notes:} Details of the calculations provided in the \hyperref[supp1:mkt]{\textit{supplementary text}}.}
\end{table}

\clearpage
\newpage

\clearpage

\phantomsection
\addcontentsline{toc}{section}{References}

\printbibliography[keyword=SI]

\end{document}